\pdfoutput=1
\documentclass{aa}  
\usepackage{amssymb}
\usepackage{amsmath}
\usepackage{amsfonts}
\usepackage{graphicx}
\usepackage{epsfig}
\usepackage{subfigure}
\usepackage[figuresright]{rotating}
\usepackage{supertabular}
\usepackage{lscape}
\usepackage{rotating}
\usepackage{txfonts}
\newcommand{\teff}{\ifmmode T_{\rm eff} \else $T_{\mathrm{eff}}$\fi}
\newcommand{\logg}{\ifmmode \log g \else $\log g$\fi}
\newcommand{\lL}{\ifmmode \log (L/L_{\odot}) \else $\log (L/L_{\odot})$\fi}
\newcommand{\mdot}{\ifmmode \dot{M} \else $\dot{M}$\fi}
\newcommand{\myr}{M$_{\odot}$ yr$^{-1}$}
\newcommand{\vsini}{\ifmmode v \sin i \else $v \sin i$\fi}
\newcommand{\vinf}{\ifmmode v_{\infty} \else $v_{\infty}$ \fi}

\newcommand{\vmac}{\ifmmode v_{\rm mac} \else $v_{\rm mac}$\fi}

\newcommand{\kms}{\ifmmode \mathrm{km~s}^{-1} \else km~s$^{-1}$\fi}
\newcommand{\msun}{\ifmmode M_{\odot} \else $M_{\odot}$\fi}
\newcommand{\zsun}{\ifmmode Z_{\odot} \else $Z_{\odot}$\fi}
\newcommand{\lsun}{\ifmmode L_{\odot} \else $L_{\odot}$\fi}
\newcommand{\rsun}{\ifmmode R_{\odot} \else $R_{\odot}$\fi}
\newcommand{\qh}{\ifmmode Q_{\rm H} \else $Q_{\rm H}$\fi}
\newcommand{\qhei}{\ifmmode Q_{\ion{He}{i}} \else $Q_{\ion{He}{i}}$\fi}
\newcommand{\mum}{\ifmmode \mu \mathrm{m} \else $\mu$m\fi}

\begin{document}
   \title{Tracing back the evolution of the candidate LBV HD\,168625 \thanks{Based in part on observations taken by {\it Herschel} satellite. {\it Herschel} is an ESA space observatory with science instruments provided by European-led Principal Investigator consortia and with important participation from NASA.} \fnmsep\thanks{Based in part on observations collected at the European Southern Observatory, in Chile.} }

   \subtitle{}

   \author{L. Mahy\inst{1}\fnmsep\thanks{F.R.S.-FNRS Postdoctoral researcher}
          \and
          D. Hutsem{\'e}kers\inst{1}\fnmsep\thanks{Senior Research Associate F.R.S.-FNRS}
          \and
          P. Royer\inst{2}
          \and
          C. Waelkens\inst{2}
          }

   \offprints{L. Mahy}

   \institute{
     Institut d'Astrophysique et de G\'eophysique, Universit\'e de Li\`ege, Quartier Agora, All\'ee du 6 Ao\^ut 19C, B-4000, Li\`ege 1, Belgium\\
     \email{mahy@astro.ulg.ac.be}
     \and 
     Instituut voor Sterrenkunde, KU Leuven, Celestijnenlaan 200D, Bus 2401, B-3001 Leuven, Belgium
   }
   
   \date{Received ...; accepted ...}
   
 
  \abstract
   {The luminous blue variable phase is a crucial transitory phase that is not clearly understood in the massive star evolution.}
   {We have obtained far-infrared {\it Herschel}/PACS imaging and spectroscopic observations of the nebula surrounding the candidate LBV HD\,168625. By combining these data with optical spectra of the central star, we want to constrain the abundances in the nebula and in the star and compare them to trace back the evolution of this object.}
   {We use the CMFGEN atmosphere code to determine the fundamental parameters and the CNO abundances of the central star whilst the abundances of the nebula are derived from the emission lines present in the {\it Herschel}/PACS spectrum. }
   {The far-infrared images show a nebula composed of an elliptical ring/torus of ejecta with a ESE-WNW axis and of a second perpendicular bipolar structure composed of empty caps/rings. We detect equatorial shells composed of dust and ionized material with different sizes when observed at different wavelengths, and bipolar caps more of less separated from the central star in H$\alpha$ and mid-IR images. This complex global structure seems to show two different inclinations: $\sim 40\degr$ for the equatorial torus and $\sim 60\degr$ for the bipolar ejections. From the {\it Herschel}/PACS spectrum, we determine nebular abundances of $\mathrm{N/H} = 4.1 \pm 0.8 \times~10^{-4}$ and $\mathrm{C/H} \sim 1.6_{-0.35}^{+1.16} \times~10^{-4}$, as well as a mass of ionized gas of $0.17\pm 0.04$\,\msun\ and a neutral hydrogen mass of about $1.0\pm 0.3$\,\msun which dominates. Analysis of the central star reveals $\teff = 14000 \pm 2000$K, $\logg = 1.74 \pm 0.05$ and $\lL = 5.58 \pm 0.11$. We derive stellar CNO abundances of about $\mathrm{N/H} = 5.0 \pm 1.5 \times~10^{-4}$, $\mathrm{C/H} = 1.4 \pm 0.5 \times~10^{-4}$ and $\mathrm{O/H} = 3.5 \pm 1.0\times~10^{-4}$, not significantly different from nebular abundances. All these measurements taken together are compatible with the evolutionary tracks of a star with an initial mass between $28$ and $33\,\msun$ and with a critical rotational rate between 0.3 and 0.4 that has lost its material during or just after the Blue Supergiant phase.}
   {}

   \keywords{circumstellar matter - Stars: massive - Stars: mass-loss - Stars: individual: HD\,168625 - Stars: individual: HD\,168607 - Stars: abundances}
   \titlerunning{}
   \authorrunning{L. Mahy et al.}
   \maketitle


\section{Introduction}
\label{sect:intro}

The luminous blue variable (LBV) stage represents a crucial and relatively short phase in the massive star evolution. The limited proportion of stars in this stage indicates that this evolutionary phase lasts only for a short time ($\sim 10^4$ years). Although a clear definition of LBVs is missing, these objects exhibit a high luminosity ($\geq 10^{5.5}L_{\odot}$) and a photometric variability, called S~Dor variability, with different timescales and different amplitudes ($\sim 0.1$~mag). Giant eruptions in which their brightness increases significantly ($> 2$~mag) for only a few years can also occur \citep[see e.g.,][for further details]{humphreys94,weis11}. This variability is responsible for the change of their spectrum. In a hot phase, the spectrum of an LBV is similar to the spectrum of a blue supergiant (BSG) whilst in a cool phase, it looks like that of an A or an F star. LBVs also exhibit episodes of high mass-loss rate ($\sim 10^{-5}-10^{-4}$~\myr). The material lost during extreme mass-loss episodes frequently produces a circumstellar nebula constituted of stellar ejecta. These nebulae indeed generally display a nitrogen overabundance, inferring a stellar origin. They can be created through giant eruptions and then shaped by wind-wind interactions. The stellar wind can become slower and denser, proceeds into a cavity carved by the massive star wind and partially fills it up with ejected material \citep{vanmarle07}.

  These nebulae have sizes that are generally estimated to about 1--2~pc. They exhibit expansion velocities that are generally between ten and a few 100\,\kms\ but can reach several 1000\,\kms\ \citep[such as the outer ejecta of $\eta$~Car,][and the references therein]{weis12}. These nebulae can also have different morphologies. \citet{weis11} yielded that about 50\% of them are bipolar, 40\% spherical and only 10\% are of irregular shape. The bipolar nebulae can display either an hourglass shape like SWB1, MN18 or Sher 25, or bipolar attached caps like WRAY\,15-751. Even though the formation of LBV nebulae is not yet fully understood, possible causes of bipolarity have been advanced, such as the rotation of the star, a density gradient in the stellar wind, or different mass-loss episodes in which the wind changes from equatorial to polar directions during the passage of the bistability jump. 

The circumstellar environment thus reveals the mass-loss history of the central star in the previous evolutionary stages. In some cases, more than one ejection can occur. \citet{jimenez10} reported multiple shells in the nebula surrounding the LBV G79.29$+$0.46, and \citet{vamvatira13} showed the presence of a second nebula around WRAY~15-751.

In the present paper, we focus on the candidate LBV HD\,168625 (or IRAS~18184--1623). This object is a very luminous hypergiant B star \citep[$M_V = -8.6$,][]{vangenderen92}. Its spectral type seems to vary from B\,2 \citep{popper40} to B\,8 \citep{morgan55} by going through B\,5.6 \citep{chentsov89} even though no dramatic light-variation has been reported in the literature in the last 40 years (\citealt{vangenderen92} and \citealt{sterken99}). Besides this lack of variation, preventing the classification of HD\,168625 as LBV, \citet{chentsov89} noticed an infrared excess at $\lambda > 5\,\mum$ that could be related to circumstellar dust. Several years later, \citet{hutsemekers94} provided clues of a high mass-loss history since they observed, for the first time, the nebula surrounding this star. They pointed out from a H$\alpha$ image that this shell consists of two regions: an inner $10\arcsec \times~13\arcsec$ elliptical ring detached from the star with a major axis oriented ESE-WNW and an outer horn-shaped (or ansa-shaped) region oriented perpendicularly to the inner region and suggesting a bipolar morphology. From a deeper picture of the H$\alpha$ nebula, \citet{nota96} detected faint filaments in this bipolar structure that extend its size to $16\arcsec \times~21\arcsec$. \citet{pasquali02} estimated the size of the nebula between $15.5\arcsec \times~23.5\arcsec$ from a 4\,\mum\ image and $31\arcsec \times~35.5\arcsec$ from an 11\,\mum\ image, both obtained with the Infrared Space Observatory (ISO). In the 4\,\mum\ image, the nebula is relatively faint in comparison with the central star whilst in the 11\,\mum\ image, the star is fainter which allows to see two bright regions of dust emission, separated by $15\arcsec$ and symmetrically located to the northwest and the southeast with respect to the central object. The torus-shaped dust emission was modeled by \citet{ohara03} who found a total dust mass of $2.5\pm0.1~10^{-3} \msun$. The question of the origin of a filamentary bipolar structure in the nebula around HD\,168625 was risen up by \citet{smith07}. A first explanation resulting from the interaction between the faster wind of the star and slower ejections was given but the possibility of other scenarios involving either the evolution of one single star or interactions between two massive stars have also been advanced. In this context, \citet{aldoretta15} detected the presence of a companion through interferometric observations. This detection has been confirmed in a very recent study of adaptative optics images by \citet{martayan16} that reveals a wide-orbit companion with a projected separation of $1.15\arcsec$. We note that this companion was previously detected by \citet{pasquali02}, who suggested already that HD\,168625 might be a wide binary system. The role of this companion could affect the shape of the nebula or the evolutionary process, but, to date, this remains conjecture. Furthermore, no hint exists that this companion is gravitationally bound to the central star.

The distance at which the star is located at was also heavily debated. \citet{hutsemekers94} and \citet{nota96} used for their respective analysis 2.2~kpc (the estimated distance of the neighboring Omega nebula M17) but the former reported that this distance could be underestimated and that HD\,168625 is probably situated behind M17 and its visual neighbor LBV HD\,168607. \citet{robberto98} determined a distance of 1.2~kpc for HD\,168625. \citet{pasquali02} estimated from the Galactic kinematic rotation model of \citet{brand93} a distance of 2.8~kpc to HD\,168625 that seems more realistic. We thus consider this last value in our analysis.

To constrain the evolutionary path of HD\,168625 and possible also to determine the influence of its putative companion during its evolution, we investigate, in this paper, the difference of CNO abundances between the nebula and the star on the basis of {\it Herschel}/PACS data and of high-resolution optical spectroscopy, respectively. We also discuss the global morphology of the nebula through a large set of H$\alpha$ and infrared images. The paper is organized as follows. In Section~\ref{sect:obs}, the observations and the data reduction procedures are presented. In Section~\ref{sec:modeling}, we model the central star by using the CMFGEN atmosphere code \citep{hillier98} to determine its fundamental parameters and the surface abundances of the elements. In Section~\ref{sec:infrared}, we present the morphology of the nebula and we analyse its {\it Herschel}/PACS spectrum. We discuss the results and constrain the evolutionary stages of HD\,168625 in Section~\ref{sec:discussion} and finally we provide the conclusions in Section~\ref{sec:conc}.


\section{Observations and data reduction}
\label{sect:obs}
\subsection{Infrared observations}
\label{subsec:infra}
\subsubsection{Imaging}

The far-infrared observations were obtained, with the PACS photometer \citep[Photodetector Array Camera Spectrometer,][]{poglitsch10} onboard the {\it Herschel} spacecraft \citep{pilbratt10}, in the framework of the Mass-loss of Evolved Stars (MESS) Guaranteed Time Key Program \citep{groenewegen11}. The imaging data were taken on March 30, 2011 which corresponds to the observational day (OD) 686 of {\it Herschel}. The scan map mode was used. In this mode, the telescope scans at constant speed, corresponding to $20\arcsec/s$ in this case, along parallel lines to cover the required area of the sky. A total of eight scans were obtained (two orthogonal scan maps at 70\,\mum\ and at 100\,\mum, four scans obtained simultaneously at 160\,\mum), over four distinct observations of 157~seconds each, carrying the following observation numbers (obsID): 1342217763, 1342217764, 1342217765 and 1342217766. The {\it Herschel} Interactive Processing Environment \citep[HIPE,][]{ott10} was used for the data reduction up to level 1. Subsequently, the Scanamorphos software \citep{roussel13} was used to further reduce and combine the data. The {\it Herschel}/PACS point spread function (PSF) full widths at half maximum (FWHMs) are $5.2\arcsec$, $7.7\arcsec$ and $12.0\arcsec$ at 70\,\mum, at 100\,\mum\ and at 160\,\mum, respectively.

We also used other space-borne images from SPITZER IRAC \citep[Infrared Array Camera,][]{fazio04} and MIPS \citep[Multiband Imaging Photometer of Spitzer,][]{rieke04}, from WISE \citep[Wide-field Infrared Survey Explorer,][]{wright10}, as well as the 11--18\,\mum\ VLT/VISIR (VLT Mid-Infrared Imager Spectrometer) images studied by \citet{umana10}, in order to better constrain the nebular environment.

\subsubsection{Spectroscopy}

The far-infrared spectrum of the nebula surrounding HD\,168625 was taken on September 25, 2011 (OD 865), with the PACS integral-field spectrometer that covers the wavelength range from 55\,\mum\ to 220\,\mum\ in two channels that operate simultaneously in the blue 55--98\,\mum\ band (B2A and B2B) and the red 102--220\,\mum\ band (R1A--R1B). It provides simultaneous imaging of a $47\arcsec \times~47\arcsec$ field-of-view, resolved in $5 \times~5$ square spatial pixels (called spaxels). The two obsIDs of these observations are 1342229740 and 1342229741. The resolving power of the spectrum is between 1000 and 3000 depending on the wavelength. HIPE was also used for the data reduction. The standard reduction steps were followed and in particular the subtraction of the background spectrum obtained through chopping and nodding. However, a special treatment was brought to the [\ion{C}{ii}]~158\,\mum\ line. This line was indeed found in absorption in the output spectrum after standard reduction. It appeared that a significant contribution to this line comes from the background spectrum. We therefore estimated manually the emission only produced by the background and we removed it properly from the nebular spectrum. The errors on the flux are thus larger for this line than for the other spectral lines detected in the spectrum.

\subsection{Visible observations}
\label{subsec:visible}

\subsubsection{Imaging}
\label{subsubsec:imaging}

We retrieved from the ESO archives the coronographic images taken in H$\alpha$ and a continuum filter ($\lambda_c = 6506\AA$, FWHM $= 35\AA$) with the SUSI camera mounted on the New Technology Telescope (NTT) at La Silla on May 23, 1995. These data were presented by \citet{nota96}. We followed the same steps for the data reduction and refer to these authors in this respect. The occulting bar has a width of $2\arcsec$ and the pixel size of the image is $0.1\arcsec$. 

\subsection{Spectroscopy}
\label{subsubsec:spectro}

We retrieved high-quality optical spectra of HD\,168625 available in the ESO archives. A first set of data was obtained with the {\'e}chelle Fiber-Fed Extended Range Optical Spectrograph (FEROS) mounted successively on the 1.52m (before 2003), then on the 2.2m telescope at La Silla (Chile). This instrument has a resolving power of 48\,000 and the detector was a 2k~$\times$~4k EEV CCD with a pixel size of $15\,\mum \times~15\,\mum$. Exposure times between 4 and 30 min were used to reach signal-to-noise ratios, measured on the 4800--4825\AA\ wavelength region, between 50 and 100, respectively. FEROS provides 39 spectral orders, covering the 3700--9200\AA\ wavelength domain. For the data reduction process, we used an improved version of the FEROS pipeline working under MIDAS environment \citep{sana06}. We combined the spectra obtained consecutively the same night to increase the signal-to-noise ratio.  \\
A second set of data was composed of three UVES (Ultraviolet and Visible Spectrograph) spectra taken at ESO Very Large Telescope (UT2, Paranal, Chile). One spectrum was taken in the dichroic mode with the DIC~1~346$+$580 and the DIC~2~437$+$860 setups and an $0.8\arcsec$ slit for all setups. The other spectra were taken with the RED 580~nm setup and an $0.8\arcsec$ slit. The resolving power of this instrument is 80\,000. The data reduction was performed with the standard reduction pipeline. \\
A third set of two spectra was acquired with the XShooter spectrograph mounted on the ESO VLT at Paranal. The data were obtained in nodding mode. Each observation covers the UBV (3\,000--5\,500\AA), the VIS (5\,500--10\,000\AA) and the NIR (10\,000--25\,000\AA) arms. The reached resolving power is about 6\,000, 10\,000 and 8\,000 in the three arms, respectively. The data were also reduced with the standard pipeline.

The 16 spectra belonging to the entire dataset were normalized by fitting polynomial functions of degree 4--5 on carefully chosen continuum windows with a width of about 200\AA. The orders were then merged to form a single spectrum. The barycentric correction was applied when it was required. The full journal of observation is provided in the Appendix.

\section{Modeling of the stellar spectrum}
\label{sec:modeling}

To determine the fundamental parameters of HD\,168625, we use the non-LTE model atmosphere code CMFGEN. This code solves the radiative-transfer equation for a spherically symmetric wind in the co-moving frame under the constraints of radiative and statistical equilibrium. The hydrostatic density structure is computed from mass conservation and the velocity structure is constructed from a pseudo photosphere structure connected to a $\beta$-velocity law of the form $v = \vinf (1- R/r)^\beta$ where \vinf is the wind terminal velocity. Our final model includes the following chemical species: \ion{H}{i}, \ion{He}{i-ii}, \ion{C}{ii-iv}, \ion{N}{i-v}, \ion{O}{i-vii}, \ion{Ca}{ii}, \ion{Na}{i}, \ion{Mg}{ii}, \ion{Ne}{ii-iii}, \ion{Ti}{ii}, \ion{Al}{ii-iii},  \ion{Cr}{ii-iii}, \ion{Si}{ii-iv}, \ion{S}{ii-v}, \ion{P}{iv-v}, \ion{Fe}{ii-vii}, and  \ion{Ni}{ii-iv} with the solar composition of \citet{grevesse10} unless otherwise stated. CMFGEN uses the super-level approach to reduce the memory requirement. In the present analysis, we include 1969 super-levels for a total of 7902 transitions. Once the emerging spectrum is generated, we convolve it first by a rotational profile determined from the Fourier transform method \citep{simondiaz07} and then we convolve the resulting spectrum by a gaussian profile to include the atmospheric macroturbulent velocity (\vmac).

\begin{figure}
\centering 
\includegraphics[width=8cm,bb=20 0 541 406,clip]{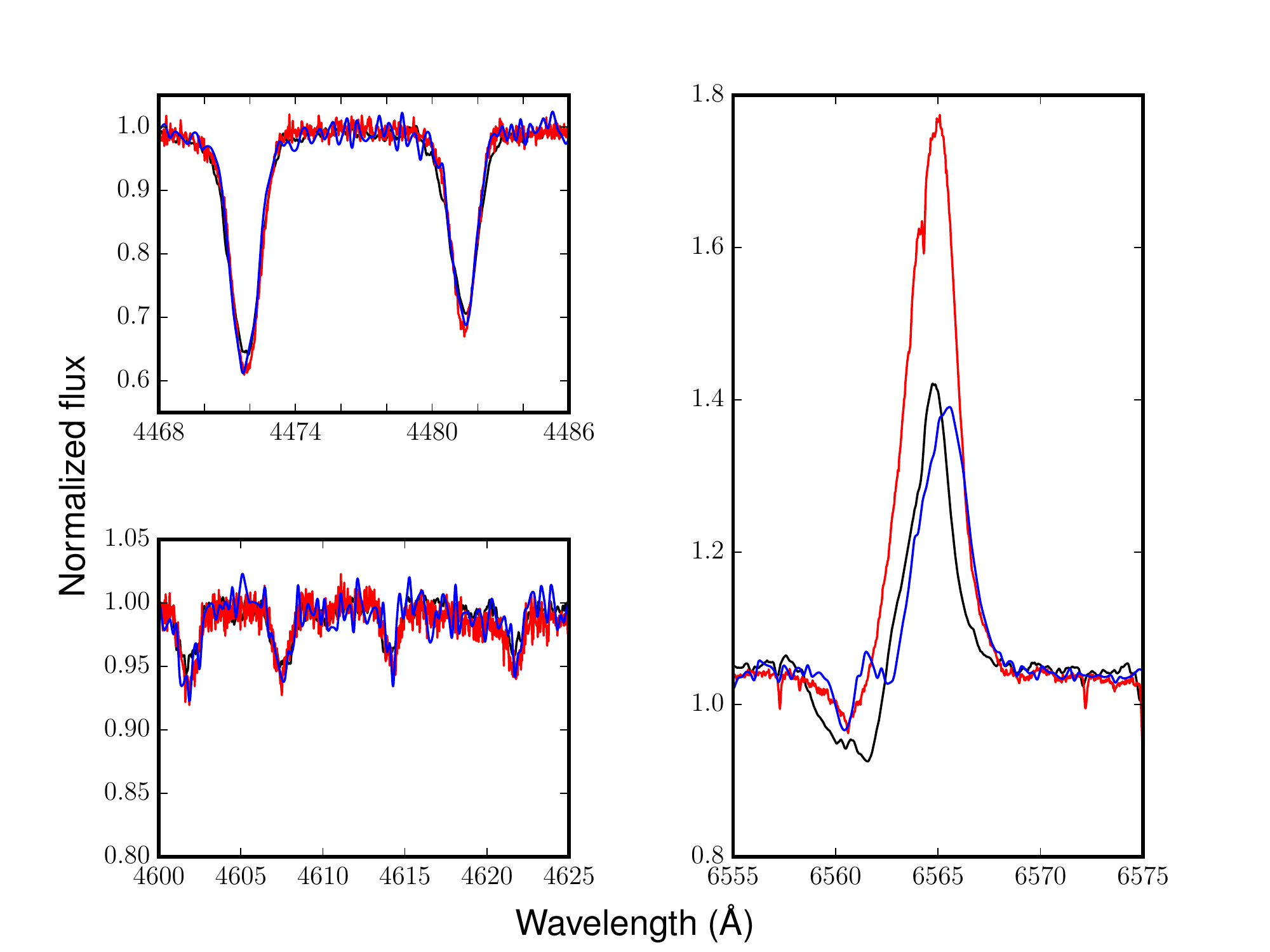}
\caption{Comparison between the FEROS spectrum taken on June 29, 2002 (black line), the UVES spectrum taken on July 10, 2001 (red line) and the XShooter spectrum taken on June 2, 2014 (blue line). The \ion{He}{i}~4471 and the \ion{Mg}{ii}~4481 lines are shown as effective temperature diagnostic, the \ion{N}{ii} lines between 4600 and 4625\AA\ as nitrogen content indicator, and the H$\alpha$ line for the stellar wind variations.}\label{fig:lines}
\end{figure}

The observed spectra do not show any double-lined spectroscopic (SB2) signature that could be related to the putative companion. We measure, from our entire dataset (FEROS, UVES, and XShooter spectra), a mean radial velocity (RV) of 19.7~\kms\ by fitting Gaussian profiles on the line profiles and a RV shift of about 15.4~\kms, but without determining a clear period. This RV shift is computed on a timescale of about 1000 days but our dataset does not allow us to be more accurate so that it must be considered as an upper limit. As for the possibly related star Sher\,25 \citep{taylor14}, the amplitude of the RV shift does not allow us to claim that HD\,168625 is a binary system, but we cannot exclude this assumption. Another difference between the optical spectra comes from H$\alpha$ profiles (see Fig.~\ref{fig:lines}). Many evolved massive stars have a H$\alpha$ profile that varies without being in a binary system because of variations of the wind parameters (\mdot, \vinf, $\beta$ and/or the clumping filling factor $f$) as a function of time. These variations do not affect the main indicators useful to determine the stellar parameters (\teff, \logg, and/or \lL) nor the abundances. All in all, we therefore consider that the optical spectrum of HD\,168625 is that of a single star. We thus determine the stellar and wind parameters from the spectrum taken on June 29, 2002. This spectrum has been chosen because of its high S/N ratio and the P-Cygni profile of the H$\alpha$ line. The investigation of the spectral variability will be carried out in a future paper. 

To estimate the main parameters of HD\,168625, we focus on the following diagnostics:
\begin{itemize}
\item Effective temperature: the estimation of \teff\ is done from the \ion{He}{i}~4471/\ion{Mg}{ii}~4481 line ratio and from ionization balances. The optical spectra of HD\,168625 allow us to only consider the \ion{Si}{iii}/\ion{Si}{ii} and \ion{Fe}{iii}/\ion{Fe}{ii} ratios.
\item Gravity: the wings of the Balmer lines are used to determine the \logg. We use H$\epsilon$, H$\delta$, H$\gamma$ and to a lesser extent H$\beta$, although H$\beta$ and H$\alpha$ are affected by the stellar wind, making them less reliable diagnostic lines for gravity.
\item Stellar luminosity: the stellar luminosity is computed from the absolute magnitude $M_V$, through the following formula: $$M_V = V -(R_V * E(B-V)) - ((5\log D)-5)$$ where $D$ is the distance to the star. We use $V=8.37$ and $(B-V)=1.41$ \citep{ducati02}. The value of $(B-V)_0$ is taken from \citet{schmidtkaler82} for a spectral type B\,6 close to that of HD\,168625. From these values, we obtain $E(B-V)=1.46$ in agreement with the value given by \citet{hutsemekers94}. Assuming $R_V=3.1$, and a distance of 2.8~kpc \citep{pasquali02}, we estimate $M_V$ to be equal to $-8.39$. If we apply the bolometric correction provided by \citet{schmidtkaler82} for a star with the same \teff, we obtain a stellar luminosity of $\lL = 5.69$. With the bolometric correction provided by \citet{martins06}, this becomes $\lL = 5.46$. For the rest of this work, we consider the average of both values, i.e., $\lL = 5.58\pm0.11$.
\item Terminal velocity: given the spectral range of our spectra, the main diagnostic line to determine \vinf\ is H$\alpha$. According to \citet{martins11}, if the line is in pure emission, \vinf\ can be measured from the line width. If the line has a P-Cygni profile, \vinf\ can be estimated from the blueward extent of the absorption part.
\item Mass-loss rate: As for \vinf, the H$\alpha$ line is the only available diagnostic line to measure \mdot. We focus on the emission part of the line to determine this quantity. We emphasize however that this value is directly bound to other parameters such as \vinf, or the clumping law.
\end{itemize}

\begin{figure*}
\centering 
\includegraphics[bb=20 0 541 406,clip]{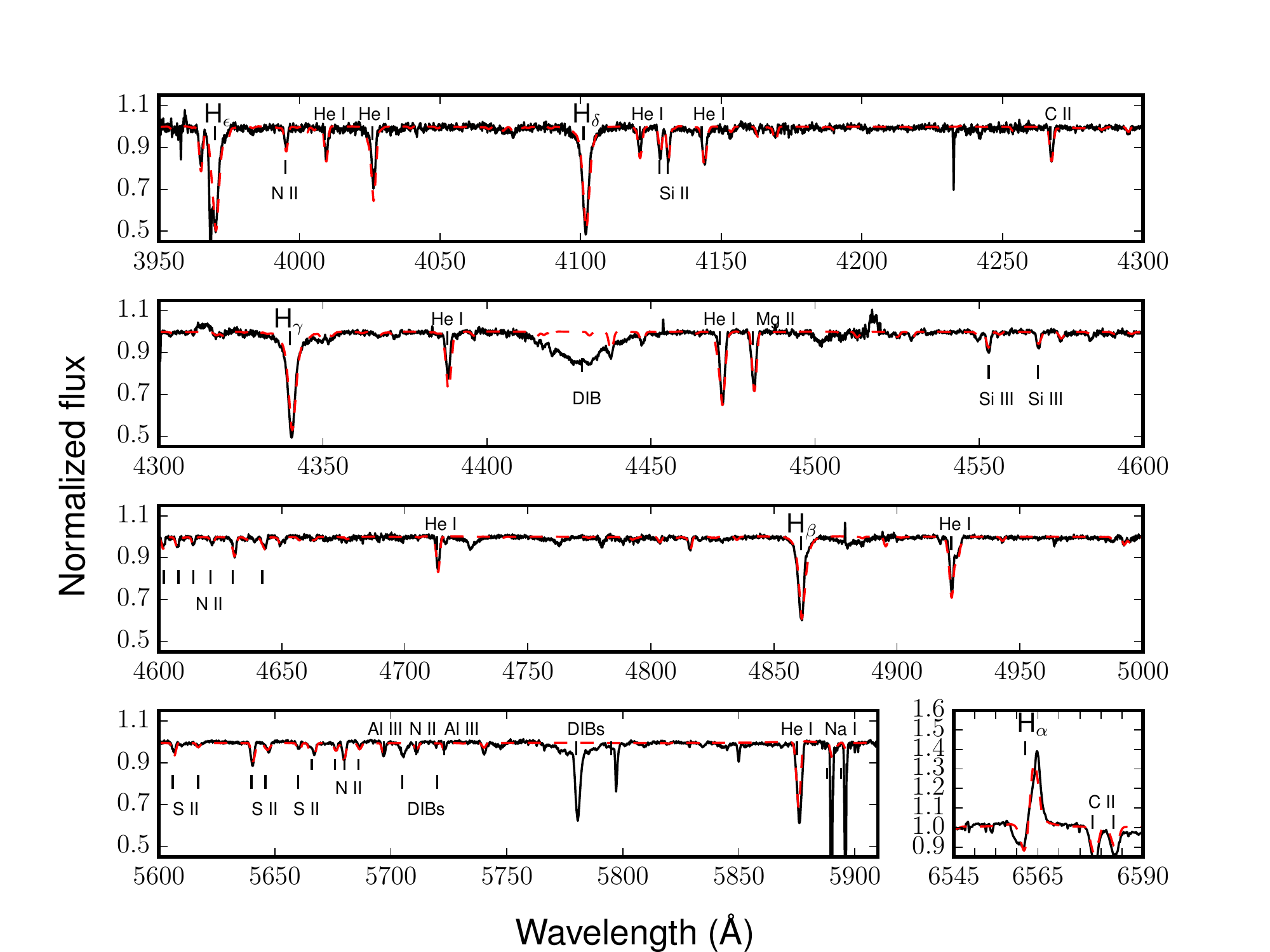}
\caption{Best-fit CMFGEN model for HD\,168625 (red line) compared to FEROS spectrum taken on June 29, 2002 (black line). The lines that are not modeled are diffuse interstellar bands (DIBs). The H$\alpha$ line has been rescaled for clarity.}\label{fig:CMFGEN}
\end{figure*}

Once the fundamental parameters are constrained, we build a grid of models changing, one by one, the He and the CNO abundances. We perform a $\chi^2$ analysis to estimate these abundances, following a method described in detail by \citet{martins15}. To briefly summarize this method, the $\chi^2$ is calculated for each chemical element by varying its abundance value. The resulting $\chi^2$ function is then renormalized so that the minimum value is equal to 1.0. We use the same lines than those quoted by these authors. The $1-\sigma$ error is set to the abundances where $\chi^2=2.0$ \citep[see][for further details]{martins15}. The fundamental parameters resulting from our CMFGEN analysis are listed in the upper panel of Table~\ref{tab:parameter} and the best-fit model is displayed in Fig.~\ref{fig:CMFGEN}.

\begin{table}
\caption{Parameters of HD\,168625 and its surrounding nebula}  
\label{tab:parameter} 
\centering            
\begin{tabular}{l c}  
\hline\hline          
\teff [K] & $14000 \pm 2000$ \\
\logg [cgs] & $1.74 \pm 0.05$ \\
$\log (L/\lsun)$& $5.58\pm 0.11$ \\
$\mdot / \sqrt{f}$ [\myr] & $1.3\times~10^{-6}$ \\
\vinf [\kms] & $350 \pm 100$  \\
$f$ & 0.3  \\
$\beta$  & 3.0 \\
He/H & $0.35\pm0.05$ \\
C/H$_{\mathrm{star}}$ & $1.4 \pm 0.5 \times~10^{-4}$ \\
N/H$_{\mathrm{star}}$ & $5.0 \pm 1.5 \times~10^{-4}$\\
O/H$_{\mathrm{star}}$ & $3.5 \pm 1.0 \times~10^{-4}$\\
\vsini\ [\kms] &  60 \\
\vmac\  [\kms] &  35  \\
\hline
$D$ [kpc] & 2.8 \\
C/H$_{\mathrm{nebula}}$ & $1.6 \pm 0.8 \times~10^{-4}$ \\
N/H$_{\mathrm{nebula}}$ & $4.1 \pm 0.8 \times~10^{-4}$ \\
\end{tabular}
\tablefoot{(He/H)$_{\odot} = 0.1$, (C/H)$_{\odot} = 2.45 \times~10^{-4}$, (N/H)$_{\odot} = 6.02 \times~10^{-5}$ and (O/H)$_{\odot} = 4.57 \times~10^{-4}$ in number.}
\end{table}

The star shows an enrichment of the helium and nitrogen abundances and a depletion in its carbon and oxygen contents. This confirms its evolved nature. Its projected rotational velocity is 60\,\kms\ which is rather slow. Its temperature range agrees with previous investigations found in the literature \citep[see, e.g.,][]{garcia-lario01} whilst its \logg\ is smaller. A comparison between the FEROS spectrum and the UVES and XShooter spectra is done in Fig.~\ref{fig:lines}. No significant effective temperature change can be seen. Furthermore, the nitrogen lines are mostly the same as a function of time. The small variations do not impact on the value of the N abundance, and are taken into account in the $1-\sigma$ error provided on this value. We emphasize that the line-profile variations are not strong enough to characterize the variability associated by an S~Dor cycle. 


\section{Analysis of the nebula}
\label{sec:infrared}

\subsection{Morphology of the nebula}
\label{subsec:morpho}

\begin{figure*}
\centering 
\includegraphics[height=5.1cm, width=6cm,bb=173 104 518 429,clip]{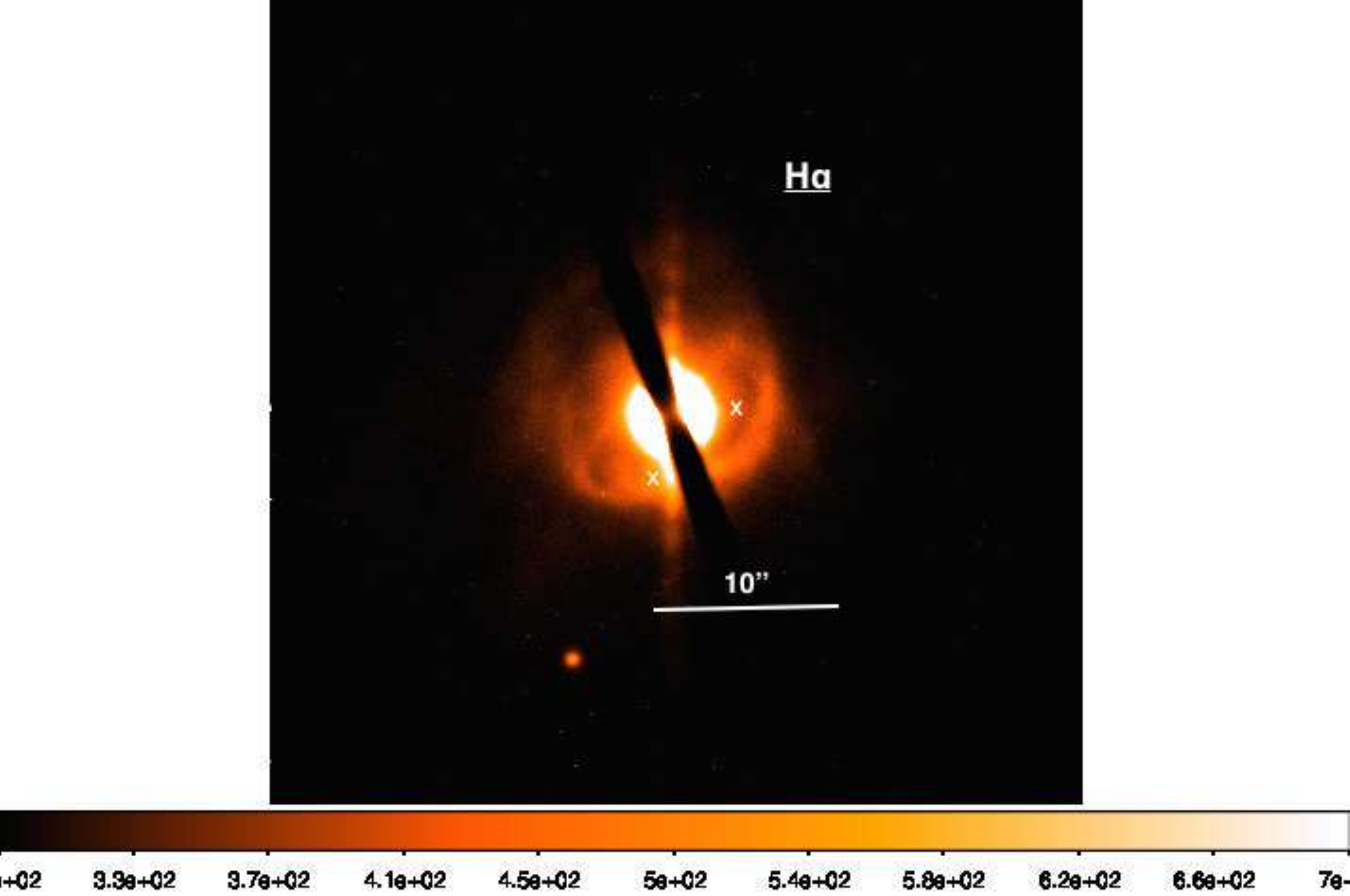}
\includegraphics[width=6cm,bb=224 73 650 434,clip]{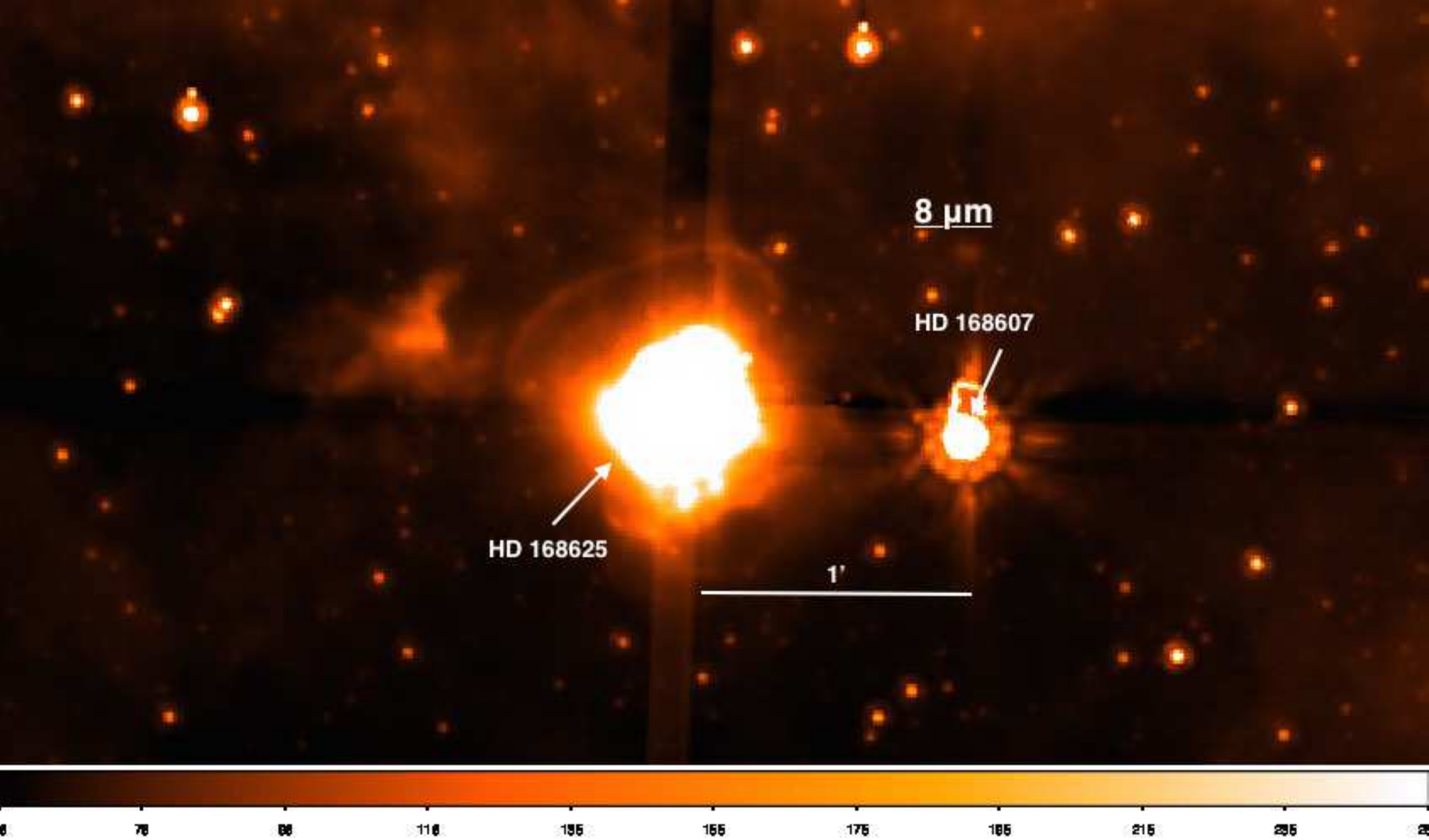}
\includegraphics[width=6cm,bb=146 65 645 493,clip]{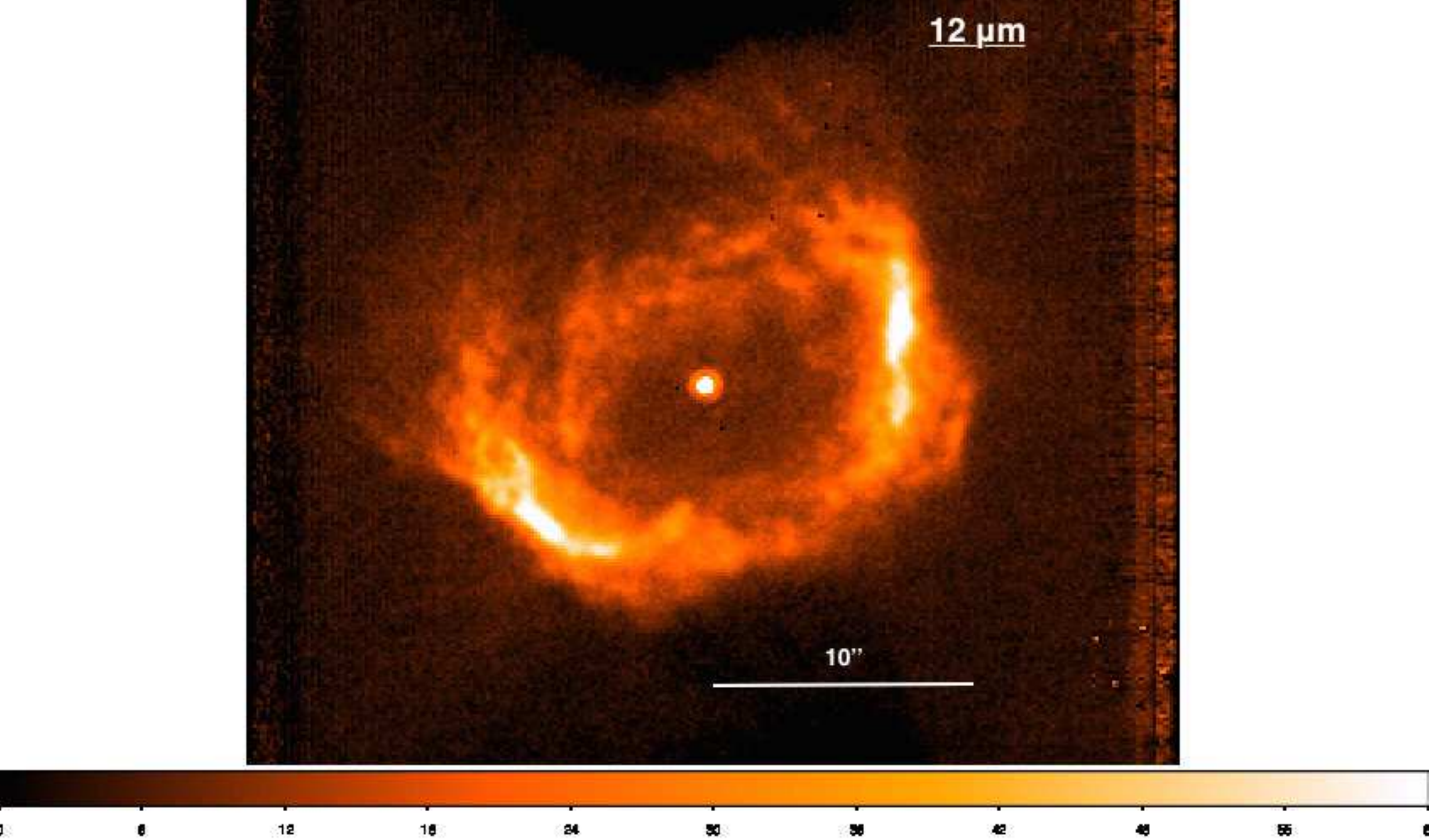}
\includegraphics[width=6cm,bb=224 110 650 404,clip]{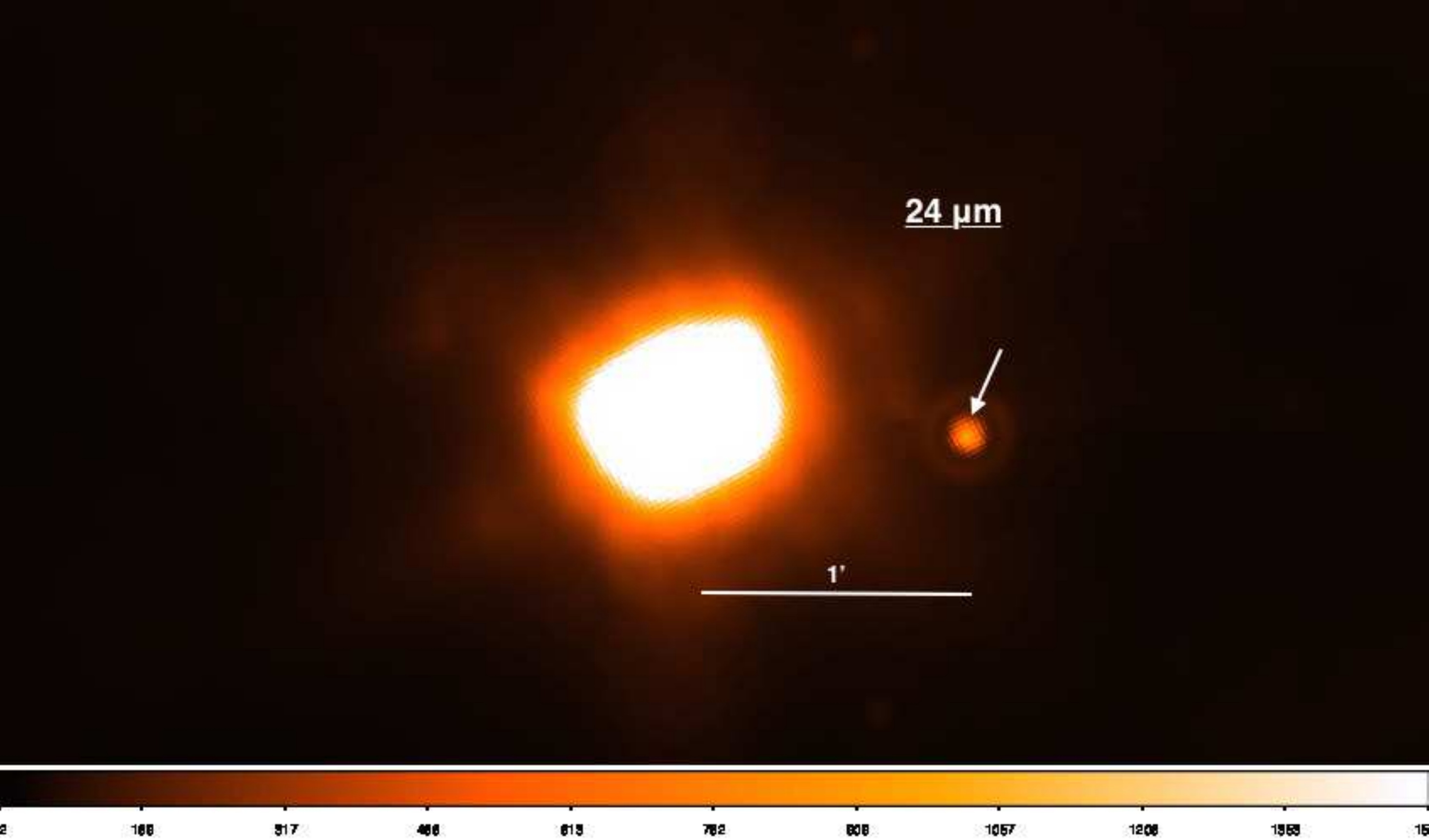}
\includegraphics[width=6cm,bb=224 110 650 404,clip]{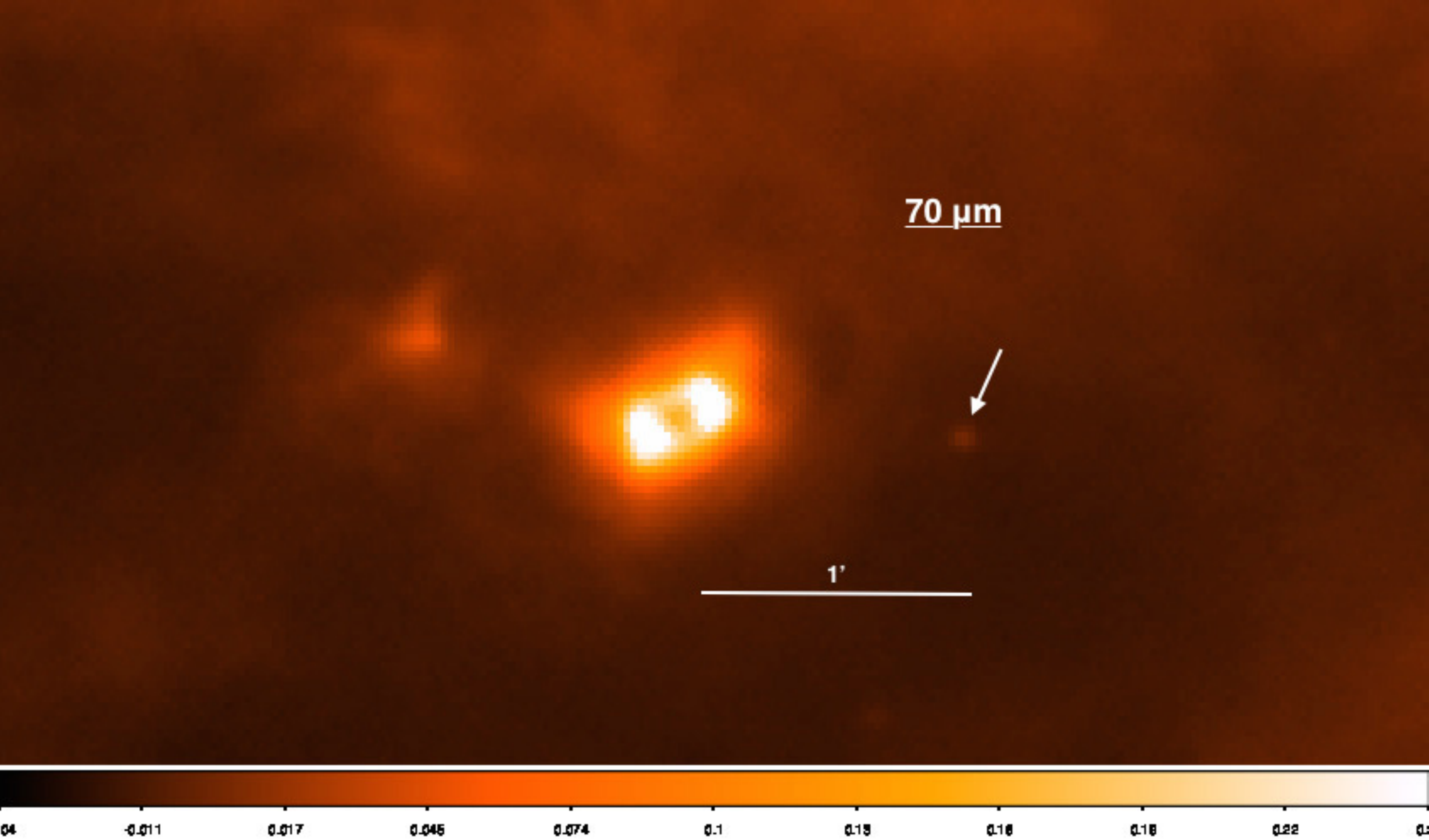}
\includegraphics[width=6cm,bb=224 110 650 404,clip]{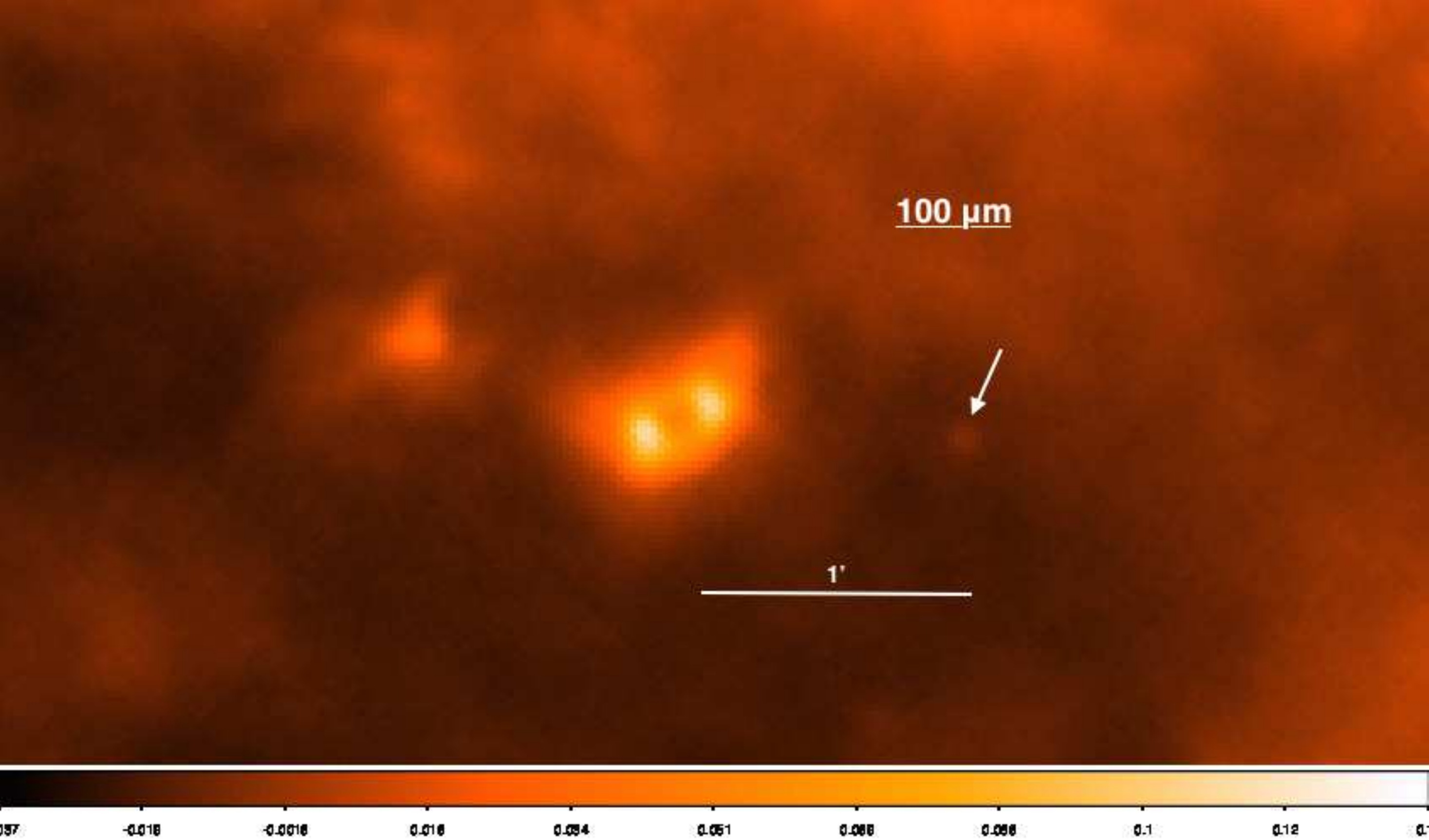}
\includegraphics[width=6cm,bb=224 110 650 404,clip]{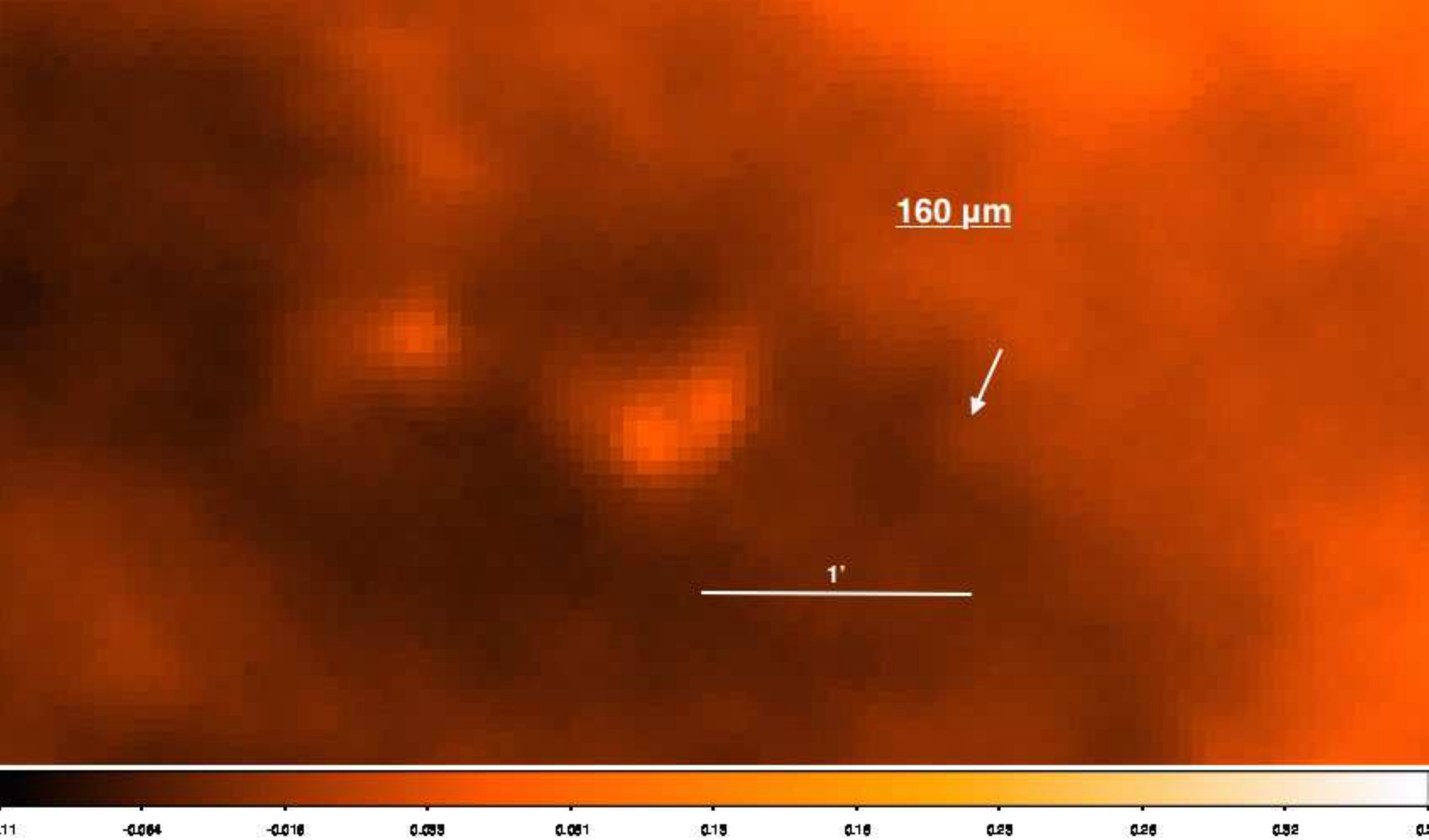}
\caption{Top: from left to right: H$\alpha$ image taken with SUSI on NTT (left),  the 8\,\mum\ SPITZER image, and the 12\,\mum\ VLT/VISIR image. Middle: the MIPSGAL 24\,\mum\ image, the PACS 70\,\mum\ image and the PACS 100\,\mum\ image from left to right. Bottom: the PACS 160\,\mum\ images. For all the images, north is up and east is left. The white arrow indicates the location of the LBV HD\,168607. In the H$\alpha$ image of \citet{nota96}, the crosses indicate the locations of the maxima of emission detected in the radio image of \citet{leitherer95}. We emphasize that the 12\,\mum\ image is from \citet{umana10} and that the inner part of the 24\,\mum\ image is saturated. }\label{fig2}
\end{figure*}

The nebula around HD\,168625 has been observed from the H$\alpha$ to 160\,\mum\ wavelength range. All these images are displayed in Fig.~\ref{fig2}. The coronographic H$\alpha$ image, taken from the data analyzed by \citet{nota96} shows the ionized nebula. This nebula is composed of two distinct structures:
\begin{itemize}
\item a bright inner elliptical torus with a major axis oriented ESE-WNW and with a size of $12\arcsec \times~16\arcsec$, corresponding to 0.16~pc~$\times$~0.22~pc at 2.8~kpc.
\item a perpendicular bipolar/biconal structure (with a major axis oriented NNE-SSW with a northern lobe clearly defined whilst the southern one is weaker. The size of this entire structure is estimated to $16\arcsec \times~21\arcsec$, i.e., $0.22$~pc~$\times\,0.29$~pc at 2.8~kpc.
\end{itemize}

The inner torus is tilted of about $40\degr$ with respect to the plane of the sky. The annular form composing the northern lobe of the bipolar/biconal structure has a size of about $8.8\arcsec \times~16.5\arcsec$, i.e., $0.12$~pc~$\times\,0.22$~pc at 2.8~kpc, which therefore infers to the bipolar/biconal structure an inclination of about $60\degr$ with respect to the plane of the sky. From the H$\alpha$ image, we also see a small cavity between the star and the inner torus. The thickness of the shell is estimated to $1.1\arcsec$ (i.e., 0.02~pc at 2.8~kpc). The inner shell is also clearly observed in the continuum image whilst the outer structure is not observed. As outlined by \citet{hutsemekers94}, it suggests that the nebula reflects the stellar luminosity by scattering due to large dust grains or shows non-equilibrium dust emission. The ionized nebula was also observed in the radio domain, at 8.64~GHz, by \citet{leitherer95}. The size of $20\arcsec$ measured from the north to the south and of $18\arcsec$ from the east to the west agrees with that estimated from the H$\alpha$ image. In the radio image, the two strongest emissions are detected at $4.5\arcsec$ to the west and at $4.5\arcsec$ to the south with respect to the central star. These locations are marked by crosses in the H$\alpha$ image of Fig.~\ref{fig2}. 

Another larger bipolar/biconal structure oriented NNE-SSW is detected in the 8\,\mum\ SPITZER image and discussed by \citet{smith07}. The northern lobe has a size of $34\arcsec \times~70\arcsec$, i.e., $0.46$~pc~$\times~0.95$~pc at 2.8~kpc. The southern one is barely visible but seems to have the same size as the northern one. The inclination of these lobes infers to the bipolar/biconal structure an angle of $60\degr$ from the plane of the sky.

Mid-infrared images have been taken by \citet{robberto98} and by \citet{umana10} to map the dust distribution in the nebula. The nebula at these wavelength is slightly larger than the ionized nebula. It is composed of an inner torus with a size of $14.5\arcsec \times~20\arcsec$, i.e., $0.19$~pc~$\times~0.27$~pc at 2.8~kpc, giving to it an inclination of about $40\degr$, in agreement with the inner torus detected in the H$\alpha$ image. We also see that the dust is highly condensed in the west-northwest and in the east-southeast part of the torus. In addition, we detect a more diffuse emission oriented perpendicularly to the dust torus, and probably related to the visible bipolar/biconal structure. We roughly estimate the size of the northern lobe to $9\arcsec \times~20\arcsec$, i.e., $0.12$~pc~$\times~0.27$~pc at 2.8~kpc. From these values, the inclination of about $60\degr$ agrees with that measured for the bipolar/biconal structure seen in the H$\alpha$ and the 8\,\mum\ SPITZER images. According to \citet{umana10}, there exists a trend in the dust emission to be more prominent in the southern part of the equatorial torus for wavelength smaller or equal to 13\,\mum\ whilst the dust emission comes preferably from the northern part of the dust torus at larger wavelengths. \citet{ohara03} interpreted this trend as a result of variations of the grain size in the different regions of the nebula or of a gradient in the dust/gas ratio. Mid-IR images also allowed \citet{skinner97} and \citet{umana10} to detect  both polycyclic aromatic hydrocarbon (PAH) and silicate features in the dust shell. The latter authors also pointed out the presence of a photodissociation region (PDR) around the ionized nebula. Furthermore, an analysis by \citet{blommaert14} of the 69\,\mum\ line observed with {\it Herschel} shows the presence of pure forsterite grains containing no iron around HD\,168625. 

The 22\,\mum\ WISE \citep[see][]{martayan16} and the MIPSGAL 24\,\mum\ images (Fig.~\ref{fig2}) should provide a relatively similar morphology of the nebula around HD\,168625. The comparison of both images shows that the inner torus-like nebula is more extended in the WISE image than in the MIPSGAL one, because of a larger width of the PSF in the WISE data. Two transversal cuts have been performed in the ESE-WNW and in the NNW-SSE directions to determine the size of the nebula. From the better quality MIPSGAL image, we estimate it to $67\arcsec \times~85\arcsec$, i.e., $0.91$~pc~$\times~1.15$~pc at 2.8~kpc, much larger than what we determine until now. At these wavelengths, the perpendicular bipolar/biconal structure is not detected. The rings detected by \citet{smith07} from the 8\,\mum\ image coincide with the border of the nebula observed at 24\,\mum. 

The far-infrared PACS images outline the cold dust distribution in the nebula around HD\,168625. These images confirm the morphology already observed from the mid-IR images. We see a torus-like nebula oriented ESE-WNW where the southeast and the northwest parts are the brightest ones. This could suggest a temperature and/or dust grain size gradient in the nebula around HD\,168625. Its size at these wavelengths is estimated to $29\arcsec \times~36\arcsec$, i.e., $0.37$~pc~$\times~0.49$~pc at 2.8~kpc. The inclination is thus about $40\degr$ as all the structures detected in the ESE-WNW axis. An "arm"-like feature is detected in the northern part (and to a lesser extent in the southern part of the nebula) on the three far-infrared images. This structure seems related to the bipolar nebula oriented along the NNE-SSW axis. However, the filamentary loops detected at 8\,\mum\ are not visible in the PACS images. Once again, it does appear that the northern lobe is the brightest one in the bipolar/biconal structure. We also note a more diffuse emission around the inner dust torus smaller than the size of the nebula estimated from the 24\,\mum\ image but with a size of $45\arcsec \times~59\arcsec$, i.e., $0.61$~pc~$\times~0.80$~pc at 2.8~kpc. The physical relation of this emission as a part of the nebula related to the star is nevertheless questionable. 

\begin{figure}
\centering 
\includegraphics[width=9cm,bb=14 55 692 502,clip]{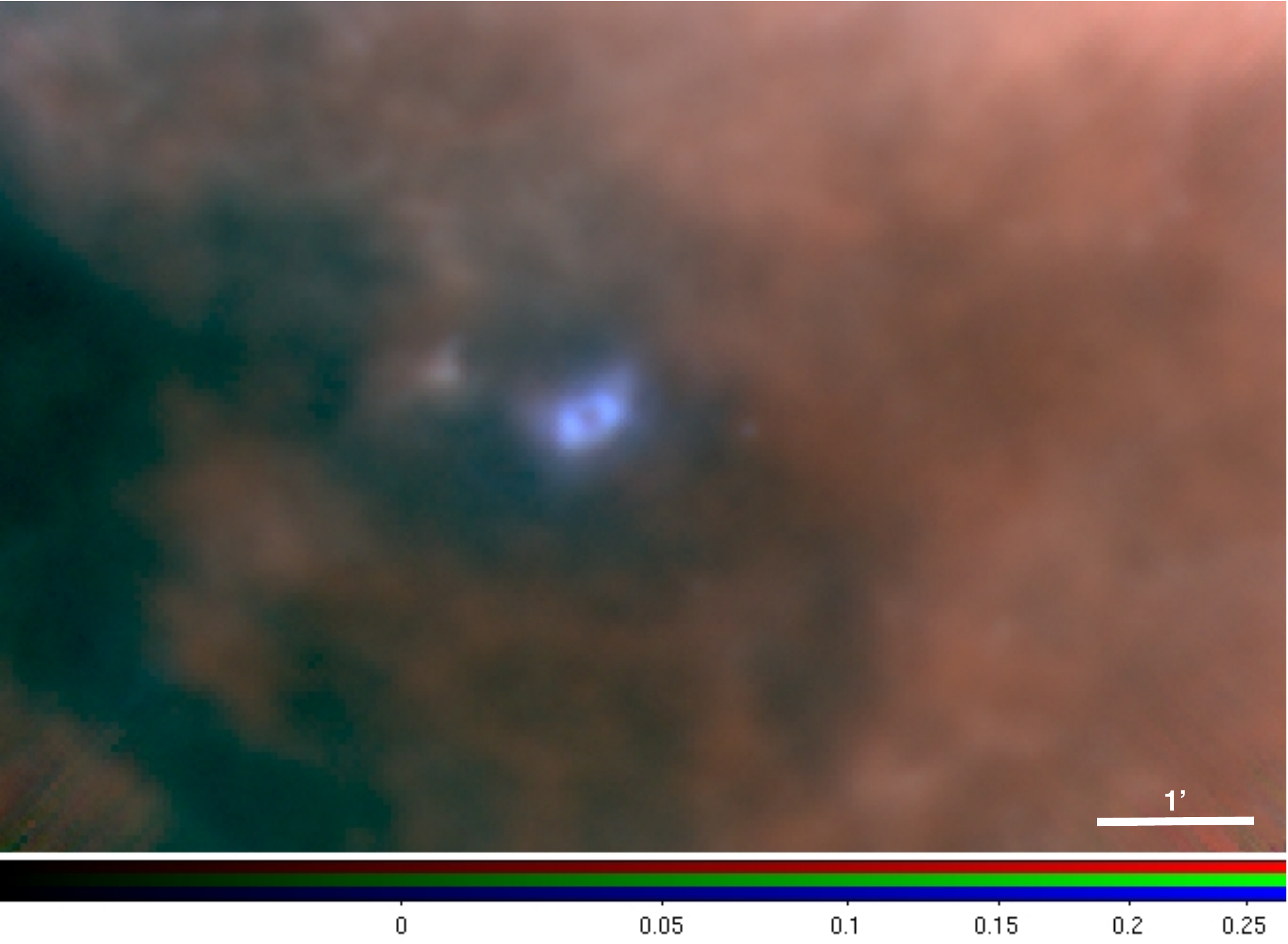}
\caption{Three-color image (70\,\mum\ in blue, 100\,\mum\ in green and 160\,\mum\ in red) of the HD\,168625 nebula. North is up and east is to the left.}\label{fig4}
\end{figure}

The three-color image (Fig.~\ref{fig4}) shows that HD\,168625 is embedded in a cavity in the interstellar medium probably formed by the star through previous mass-loss episodes. We also see that HD\,168607 that is located at $1\arcmin$ of HD\,168625 does not show any associated nebula. No cavity seems to exist around this confirmed LBV from the PACS images whilst a clear cavity was observed by \citet{hutsemekers94} in their H$\alpha$ image.

The global morphology of the nebula around HD\,168625 is thus extremely complex. To summarize, it is composed of torus-like shells visible at different wavelengths and oriented in the ESE-WNW direction. These structures seems to be tilted of $40\degr$ with respect to the plane of the sky. In addition, a perpendicular bipolar/biconal structure, oriented in the NNW-SSE direction, is also detected in the H$\alpha$, 8\,\mum, and 13\,\mum\ images. This structure seems to have an inclination of $60\degr$ with respect to the plane of the sky. Furthermore, the northern lobe is brighter than the southern one.

\subsection{The Infrared nebular spectrum}
\label{subsec:spectrum}

\begin{figure}
\centering 
\includegraphics[width=8cm,bb=0 0 530 424,clip]{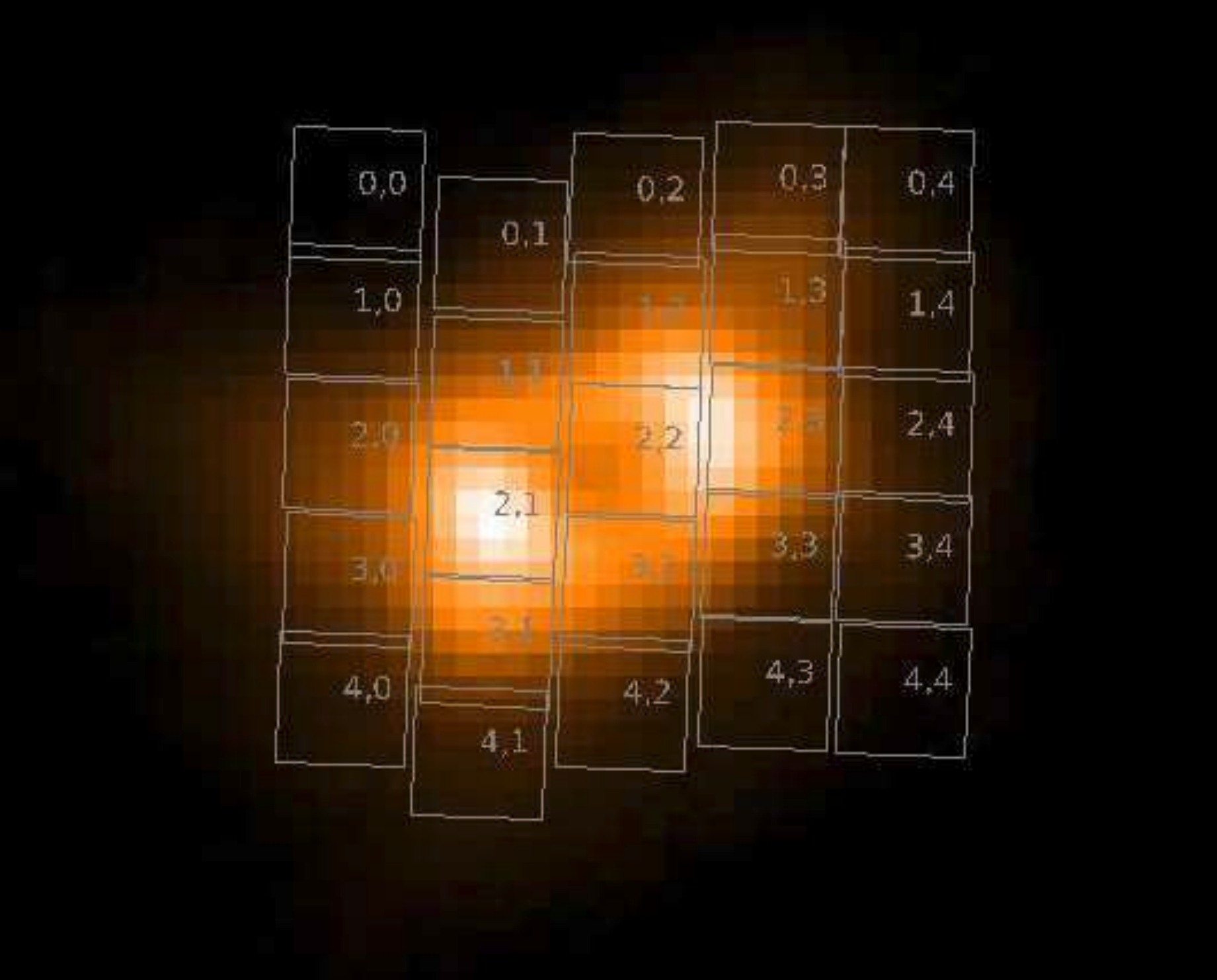}
\caption{Footprint of the PACS spectral field-of-view on the image at 70\,\mum\ of the nebula. The spaxels are represented in gray. North is up and east is left.}\label{fig9}
\end{figure}

Figure~\ref{fig9} illustrates the footprint of the PACS spectral field-of-view on the image of the nebula at 70\,\mum. This field-of-view is composed of 25 ($5\times~5$) spaxels, each corresponding to a different area of the nebula. The nebula seems to be covered by the entire spectral field-of-view but to be sure that no flux is lost by the detector, we compare the flux of the spectral continuum to the photometric flux densities measured from the 70\,\mum, the 100\,\mum\ and the 160\,\mum\ images (for further details on the method, see \citealt{vamvatira13}). For HD\,168625, a good agreement exists and no correction factor must be applied. The analysis of the dust Spectral Energy Distribution including PACS data will be discussed in another paper although the results are relatively close to those provided by \citet{ohara03}.

\begin{figure*}
\centering 
\includegraphics[bb=5 5 537 408,clip]{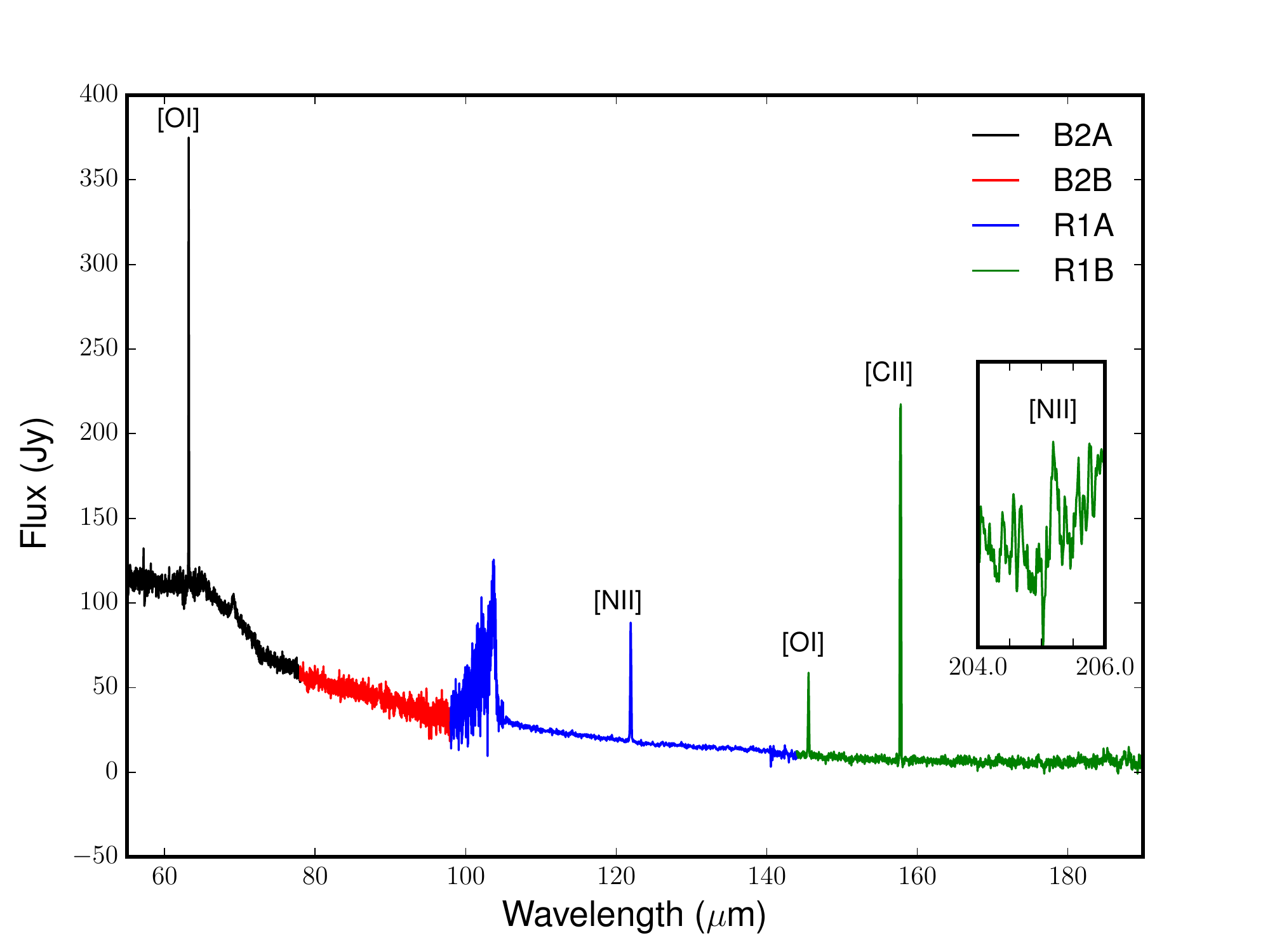}
\caption{PACS spectrum of the HD\,168625 nebula, integrated over the 25 spaxels. The different bands are indicated with different colors.The continuum and the line at 205\,\mum\ are not calibrated, the inset is expressed in arbitrary units. The spectrum between 95 and 102\,\mum is affected by an instrumental artefact that makes the spectrum scientifically unusable in this wavelength domain. }\label{fig10}
\end{figure*}

The integrated spectrum of the nebula over the 25 spaxels exhibits the [\ion{O}{i}]~63\,\mum, [\ion{N}{ii}]~122\,\mum, [\ion{O}{i}]~146\,\mum, [\ion{C}{ii}]~158\,\mum , and [\ion{N}{ii}]~205\,\mum\ lines (Fig.~\ref{fig10}). Furthermore, this spectrum also contains the band of forsterite at 69\,\mum\ that was discussed by \citet{blommaert14}. Given the spectral type of the central star, it is obvious that the spectrum of the surrounding nebula cannot display any higher ionization lines such as [\ion{N}{iii}]~57\,\mum\ or [\ion{O}{iii}]~88\,\mum. We emphasize that the [\ion{N}{ii}]~205\,\mum\ line has a problematic calibration in PACS. This issue has already been mentioned by \citet{vamvatira13,vamvatira15}. To summarize, the flux measured for this line is incorrect because of a light leak that superimposes a significant amount of light from 102.5\,\mum\ at that wavelength. The relative spectral response function (RSRF) used to reduce the data suffers from the same light leak. Consequently, when the RSRF is applied during the data reduction process, the signal at wavelengths larger or equal to 190\,\mum\ is divided by a too high number. Therefore, the flux of the [\ion{N}{ii}]~205\,\mum\ line must be corrected by a factor of 4.2. An error of 25\% is assumed for the corrected final flux of the [\ion{N}{ii}]~205\,\mum\ line.

To measure the intensies of the lines from the whole nebula, we perform a gaussian fit on the different line profiles of the integrated spectrum. The measurements are listed in Table~\ref{tab:fluxes}. We also do the same measurements on all lines present in the different spaxels to estimate the flux distribution across the nebula. As we already mentioned, the errors on the flux measurements of the [\ion{C}{ii}]~158\,\mum\ line are larger than for the other lines. The flux measurements on the different spaxels are given in the Appendix.

\begin{table}
\caption{Line fluxes from the nebula around HD\,168625}             
\label{tab:fluxes}      
\centering                          
\begin{tabular}{l c c}        
\hline\hline                 
Line & Flux  & Error         \\    
     & [$10^{-15}$ W m$^{-2}$]& [$10^{-15}$ W m$^{-2}$]   \\
\hline                        
    $[\ion{O}{i}]\,63\,\mum$   & 12.2 &  2.2 \\
    $[\ion{N}{ii}]\,122\,\mum$ & 2.1 & 0.8\\
    $[\ion{O}{i}]\,146\,\mum$  &  1.1 & 0.4\\
    $[\ion{C}{ii}]\,158\,\mum$ &  4.5 & 1.4\\
    $[\ion{N}{ii}]\,205\,\mum$ &  $0.09 \times~4.2^{a}$ & 0.1\\ 

\hline                                   
\end{tabular}
\tablefoot{$^a$: PACS calibration correction}
\end{table}

\subsection{Electron density and abundances}
\label{subsec:electron}

 \citet{nota96} determined an upper limit to the electronic temperature $T_e$ of 7000~K as well as the electron density distribution $n_e$ across the nebula. The density measured at $3\arcsec$ to the north shows a distribution between 200 and 1600~cm$^{-3}$ whilst $n_e$ reaches between 400 and 3000~cm$^{-3}$ at $2\arcsec$ to the south. These authors inferred a mean value of 1000~cm$^{-3}$ from all their measurements.

In the optical domain, the electron density is computed from the ratio [\ion{S}{ii}]~6717\AA/[\ion{S}{ii}]~6731\AA. \citet{hutsemekers94} estimated this ratio to about 0.9., giving an electron density of $n_e = 630\pm 300$~cm$^{-3}$ for the whole nebula around HD\,168625. In the infrared waveband, the ratio between [\ion{N}{ii}]~122\,\mum\ and [\ion{N}{ii}]~205\,\mum\ is a good indicator to compute the electron density of the nebula. Considering the values provided in Table~\ref{tab:fluxes}, we compute a ratio [\ion{N}{ii}]~122/[\ion{N}{ii}]~205~$=5.7 \pm 2.9$ for the whole nebula. By using the nebular/IRAF package, and by assuming a $T_e$ of 7000~K, we determine an electron density of $n_e = 350 \pm 270$~cm$^{-3}$. We consider this estimate of the electron density in the following calculations because the electron density is best determined when it is similar to the critical density of the lines whose ratio is used as diagnostic \citep{rubin94}.

In the literature, we find many different estimations of the H$\alpha$ flux from the nebula around HD\,168625. \citet{hutsemekers94} estimated a reddened flux of $4.6 \times~10^{-13}$~erg~s$^{-1}$~cm$^{-2}$ and a dereddened flux of $1.2 \times~10^{-11}$~erg~s$^{-1}$~cm$^{-2}$ given a $E(B-V) = 1.46$. \citet{nota96} provided values of $1.8 \times~10^{-12}$~erg~s$^{-1}$~cm$^{-2}$ and of $3.7 \times~10^{-10}$~erg~s$^{-1}$~cm$^{-2}$ for the reddened and the dereddened fluxes, respectively, by assuming $E(B-V)=1.86$. Finally, \citet{pasquali02} estimated a reddened and a dereddened flux of $4.0 \times~10^{-12}$~erg~s$^{-1}$~cm$^{-2}$ and $9.6 \times~10^{-10}$~erg~s$^{-1}$~cm$^{-2}$, respectively. It is thus clear that the final value of the H$\alpha$ flux depends on the extinction (different from one study to another) and on the contamination by dust scattering of the stellar continuum and H$\alpha$ emission that is difficult to correct for. Therefore, to be more accurate on the H$\alpha$ flux, we use the radio flux of $14 \pm 2$~mJy measured at 8.64~GHz by \citet{leitherer95} and we convert it into H$\alpha$ flux using Eq.~1 of \citet{vamvatira16}. We obtain $F_0($H$\alpha) = 1.84 \pm 0.26 \times~10^{-11}$~erg~s$^{-1}$~cm$^{-2}$. We then apply the nebular/IRAF package to determine the N/H abundance ratio of the nebula. We assume a case-B recombination \citep{draine11}, with $T_e = 7000$~K and with $n_e = 350$~cm$^{-3}$, and we derive a nitrogen abundance of N/H~$=4.1 \pm 2.1 \times~10^{-4}$ by number, i.e., $12 + \log (N/H) = 8.61 \pm 0.29$. This value confirms the enrichment of the nebula in nitrogen and thus that the ejecta has a stellar origin. Our value computed using $n_e = 350$\,cm$^{-3}$ is larger than the average value of \citet{nota96} of $12 + \log (N/H) = 8.04$ computed using $n_e = 700$ up to 2300\,cm$^{-3}$ for different regions of the nebula. The H$\alpha$ flux, different in both analyses, also explains the different values of the nitrogen abundance in the nebula.

\subsection{Ionizing flux}
\label{subsec:ionflux}

From the H$\alpha$ flux, we calculate the rate of emission of hydrogen-ionizing photons, $Q_0$ using the equations provided by \citet{vamvatira13}. For the nebula around HD\,168625, we estimate the rate of emission of hydrogen-ionizing photons to $Q_0 = 1.09 \pm 0.16 \times~10^{46}$ photons s$^{-1}$ with an electron temperature of $T_e = 7000$~K and a distance of 2.8~kpc. From the CMFGEN model, we determine a rate of emission of hydrogen-ionizing photons from the star of $Q_0 = 6.9\pm 4.0 \times~10^{45}$ photons s$^{-1}$, which is in agreement, within the uncertainties, with the value obtained from the H$\alpha$ flux.

The radius of the Str{\"o}mgren sphere is calculated using $Q_0$ determined from the H$\alpha$ emission. We obtain $R_S = 0.13 \pm 0.08$~pc ($\sim 9.6\arcsec \pm 6.5\arcsec$ at 2.8~kpc), assuming that the star is not hot enough to ionize the helium, that the ionized gas fills in the whole volume of the nebula and $T_e =7000$~K. To have a radius of the Str{\"o}mgren sphere equals to the radius of the ionized nebula detected from the H$\alpha$ image, we must assume that the ionized gas fills in only about 20\% of the entire volume of the nebula. This also leads to the conclusion that the H$\alpha$ nebula is likely ionization bounded. 

The mass of the ionized gas can be estimated from the H$\alpha$ flux. We consider the formula given by \citet{vamvatira13} to determine this mass. Since the star has $\teff = 14\,000$~K, and its spectrum does not show ionized helium, the mass of the ionized nebula is computed to be $M_{i_{(H\alpha)}} = 0.17 \pm 0.04\,\msun$.

\subsection{Photodissociation region}
\label{subsec:PDR}

The [\ion{O}{i}]~63\,\mum, [\ion{O}{i}]~146\,\mum\ and  [\ion{C}{ii}]~158\,\mum\ lines probably indicate the presence of a PDR around HD\,168625. This region usually surrounds the ionized region of the nebula and is composed of neutral gas. These lines can also reveal the existence of shocks, that would be the result of interactions between the fast stellar wind and the slow expanding remnant of a previous evolutionary phase. The calculated ratios between, on the one hand, the [\ion{O}{i}]~63\,\mum\ and the [\ion{O}{i}]~146\,\mum\ and, on the other hand, between the [\ion{O}{i}]~63\,\mum\ and the [\ion{C}{ii}]~158\,\mum\ are in agreement with the PDR models of \citet{kaufman99} and not with the shock models of \citet{hollenback89}. The [\ion{O}{i}]~63\,\mum/[\ion{C}{ii}]~158\,\mum\ ratio is, furthermore, a reliable indicator of the true origin of the lines. This value is indeed found smaller than 10 in PDRs \citep{tielens85} whilst it is larger for shocks. We can thus conclude that a PDR and not a shock is responsible for the presence of those lines in the spectrum of the nebula around HD\,168625. We note that this conclusion is unchanged after correcting the [\ion{C}{ii}] for the contribution of the \ion{H}{ii} region as detailed below. \citet{umana10} already detected the PDR in the nebula of HD\,168625 through spectral features in the [11--18]\,\mum\ range indicating the presence of polycyclic aromatic hydrocarbons (PAHs). To determine the physical conditions in the PDR, we use the three infrared lines quoted above. We must however subtract any possible contribution to the observed lines of the \ion{H}{ii} region. 

Neutral oxygen can be found only in neutral regions. Therefore, no doubt exists that the [\ion{O}{i}] lines arise exclusively from the PDR \citep{malhorta01}. Neutral atomic carbon has however an ionization potential (11.26~$eV$) lower than that of hydrogen, and can thus be found in the PDRs as well as in the \ion{H}{ii} regions.

\begin{figure*}
\sidecaption
\includegraphics[width=12cm,bb=25 5 538 402,clip]{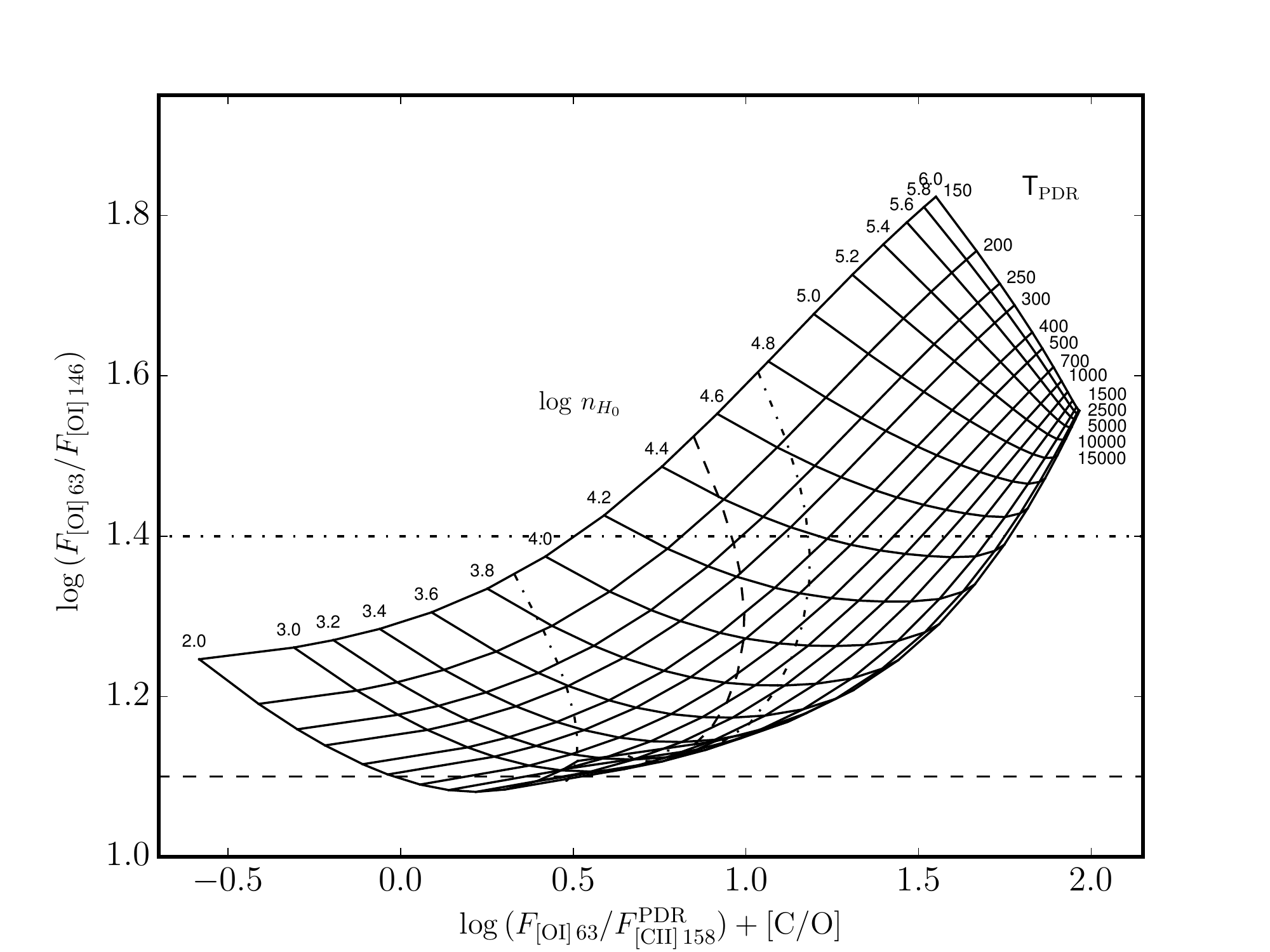}
\caption{Temperature-density PDR diagnostic diagram. The grid of flux ratios $F_{[\ion{O}{i}]~63}/F_{[\ion{O}{i}]~146}$ versus $F_{[\ion{O}{i}]~63}/F^{\mathrm{PDR}}_{[\ion{C}{ii}]~158}$ was calculated by solving the level population equations for a range of temperatures and densities. The dashed lines corresponds to the pressure equilibrium constraint between the \ion{H}{ii} region and the PDR, with the corresponding errors (dash-dot lines). The horizontal dashed lines correspond to the observed ratio with the errors \citep[see][for more details]{vamvatira13}.} \label{fig:PDRdiag}
\end{figure*}

The C/O abundance ratio can be estimated based on the PDR line fluxes and following a method described by \citet{vamvatira13} that was used to separate the contributions of the PDR and of the \ion{H}{ii} region to the flux of [\ion{C}{ii}]~158\,\mum. In the ionized gas region, the ratio of fractional ionization is given by
\begin{equation}
  \frac{<\mathrm{C}^+>}{<\mathrm{N}^+>} = \frac{F^{\ion{H}{ii}}_{[\ion{C}{ii}]~158}/\epsilon_{[\ion{C}{ii}]~158}}{F_{[\ion{N}{ii}]~122}/\epsilon_{[\ion{N}{ii}]~122}}
\end{equation}
where we define $F^{\ion{H}{ii}}_{[\ion{C}{ii}]~158}=\alpha \, F_{[\ion{C}{ii}]~158}$ with $ F_{[\ion{C}{ii}]~158}$ being the total flux of the [\ion{C}{ii}]~158\,\mum\ line and $\alpha$ as a factor to be determined. Assuming that $<\mathrm{C}^+>/<\mathrm{N}^+> = $ C/N and calculating the emissivities using the nebular/IRAF package with $n_e = 350$~cm$^{-3}$ and $T_e = 7000$~K, we find
\begin{equation}
  \frac{F^{\ion{H}{ii}}_{[\ion{C}{ii}]~158}}{F_{[\ion{N}{ii}]~122}} = [0.37 \pm 0.24] \frac{C}{N}.
\end{equation}
The N/O abundance ratio is not known in the literature, even though \citet{nota96} assumed that it is larger than $3 \times~$(N/O)$_{\odot}$. Since the nebula characterizes the star at an earlier epoch, it seems unlikely that the N/O ratio of the nebula is larger than the surface N/O ratio measured from the star. Therefore, we define an upper limit of N/O equal to 1.4.
We first assume that N/O is equal to 1. We obtain
\begin{equation}
  \log \alpha = \log~\mathrm{C/O} - 0.767,
\end{equation}
using the measured $F_{[\ion{C}{ii}]}/F_{[\ion{N}{ii}]}$ ratio and $\mathrm{C/O} = \mathrm{C/N} \times \mathrm{N/O}$.
Assuming that there is a pressure equilibrium between the ionized gas region and the PDR, we have
\begin{equation}
n_{H_0}~k~T_{\mathrm{PDR}} \cong 2~n_e~k~T_e = 4.9 \pm 3.7 \times~10^6 \mathrm{cm}^{-3} \mathrm{K}
\end{equation}
that is used to define a locus of possible values in the diagram of Fig.~\ref{fig:PDRdiag}.

The structure of the PDR is described by the density of the atomic hydrogen, $n_{H_O}$, and the incident FUV radiation field, $G_0$, that can be computed using the following equation \citep{tielen05}:
\begin{equation}
  G_0 = 625 \frac{L_{\star}\, \chi}{4\pi R^2}
\end{equation}
where $L_{\star}$ is the stellar luminosity, $\chi$ is the fraction of this luminosity above $6~eV$, equal in the case of HD\,168625 to $0.4\pm 0.05$ \citep{young02} and $R$ is the radius of the ionized gas region surrounded by the PDR (i.e., 0.22~pc). $G_0$ is expressed in terms of the average interstellar radiation field that corresponds to an unidirectional radiation field of $1.6 \times~10^{-3}$~erg~s$^{-1}$~cm$^{-2}$. $G_O$ is found to be about $6.3 \times~10^{4}$ for the PDR around HD\,168625. This value coupled with the figure~1 of \citet{kaufman99} provides a temperature of the PDR of about 1600~K. This gives $\log n_{H_0} = 3.48$ (Eq.~4). From these values and Fig.~\ref{fig:PDRdiag}, we derive $\log (F_{[\ion{O}{i}]~63}/F^{\mathrm{PDR}}_{[\ion{C}{ii}]~158}) + [\mathrm{C/O}] = 0.5 $, where by definition $[\mathrm{C/O}] = \log(\mathrm{C/O}) - \log(\mathrm{C/O})_{\odot}$. Using the values of the flux densities from Table~\ref{tab:fluxes} and $(\mathrm{C/O})_{\odot} = 0.5$ \citep{grevesse10}, we solve the equations to obtain $\alpha = 0.09$ and $\mathrm{C/O} = 0.53$. By considering N/H~$ = 4.1\times~10^{-4}$ (Sect.~\ref{sec:modeling}), we have C/H~$= 2.2\times~10^{-4}$. Assuming different values of N/O going from $\mathrm{N/O}=0.4$ \citep[see][and the references therein]{smith07} to $\mathrm{N/O}=10$ through $\mathrm{N/O}=1.4$ which is the N/O ratio determined for the central star, the resulting values of $\alpha$, C/O, and C/H are listed in Table~\ref{tablePDR}. We can discard N/O values that provide us with C/H ratios larger than the solar value. Moreover, since (N/H)$_\mathrm{neb}$~=~(N/H)$_\mathrm{star}$, it is reasonable to have (C/H)$_\mathrm{neb} \sim\,$(C/H)$_\mathrm{star}$. We therefore conclude that N/O is between 0.8 and 1.4 and we adopt $\mathrm{C/H} = 1.6_{-0.35}^{+1.16} \times 10^{-4}$ in the nebula.

\begin{table}
\caption{Nebular abundance ratios and mass from PDR modeling}             
\label{tablePDR}      
\centering                          
\begin{tabular}{l c c c c c}        
\hline\hline                 
N/O   & 0.4  & 1.0 & 1.4 & 5.0 & 10.0         \\    
\hline                        
$\alpha$               & 0.04 & 0.09 & 0.12 & 0.33 & 0.50 \\
C/O                    & 0.56 & 0.53 & 0.51 & 0.39 & 0.29 \\
C/H [$\times~10^{-4}$] & 5.7  & 2.2  & 1.5  & 0.3  & 0.1  \\
$M_H$ [\msun]          & 0.36 & 0.90 & 1.25 & 4.48 & 8.96 \\
\hline                                   
\end{tabular}
\end{table}

The total mass of the hydrogen in the PDR can be estimated using the [\ion{C}{ii}]~158\,\mum\ line flux, derived for the PDR and the equation given by \citet{vamvatira13}. For $\mathrm{C/H} = 1.6 \times 10^{-4}$, we estimate a mean value for the neutral hydrogen mass to $M_H = 1.0 \pm 0.3\,\msun$. Most of the nebular mass is thus contained in the neutral gas and not in the ionized gas.


\section{Discussion}
\label{sec:discussion}

\subsection{Evolutionary status}

The LBV phase is still poorly understood. How and when this evolutionary phase occurs are questions that must still be debated. Currently, more and more studies aim at showing that LBVs could be binary systems. Although the term "binary" must be clarified, $\eta$\,Car \citep{daminelli97} and MWC\,314 \citep{lobel13} were clearly reported as binaries whilst one has detected for HR\,Car \citep{rivinius15}, and HD\,168625 \citep{pasquali02,martayan16} close objects on wide orbits that could be related to companions. If a physical orbit is confirmed for those possible companions, we can wonder whether the LBV phase could be related to binary interactions or whether the current single star evolutionary tracks can explain this evolutionary stage, as it seems to be the case for WRAY\,15-751 \citep{vamvatira13} and for AG\,Car \citep{vamvatira15}.

\begin{figure}
\centering 
\includegraphics[width=8cm,bb=25 5 538 402,clip]{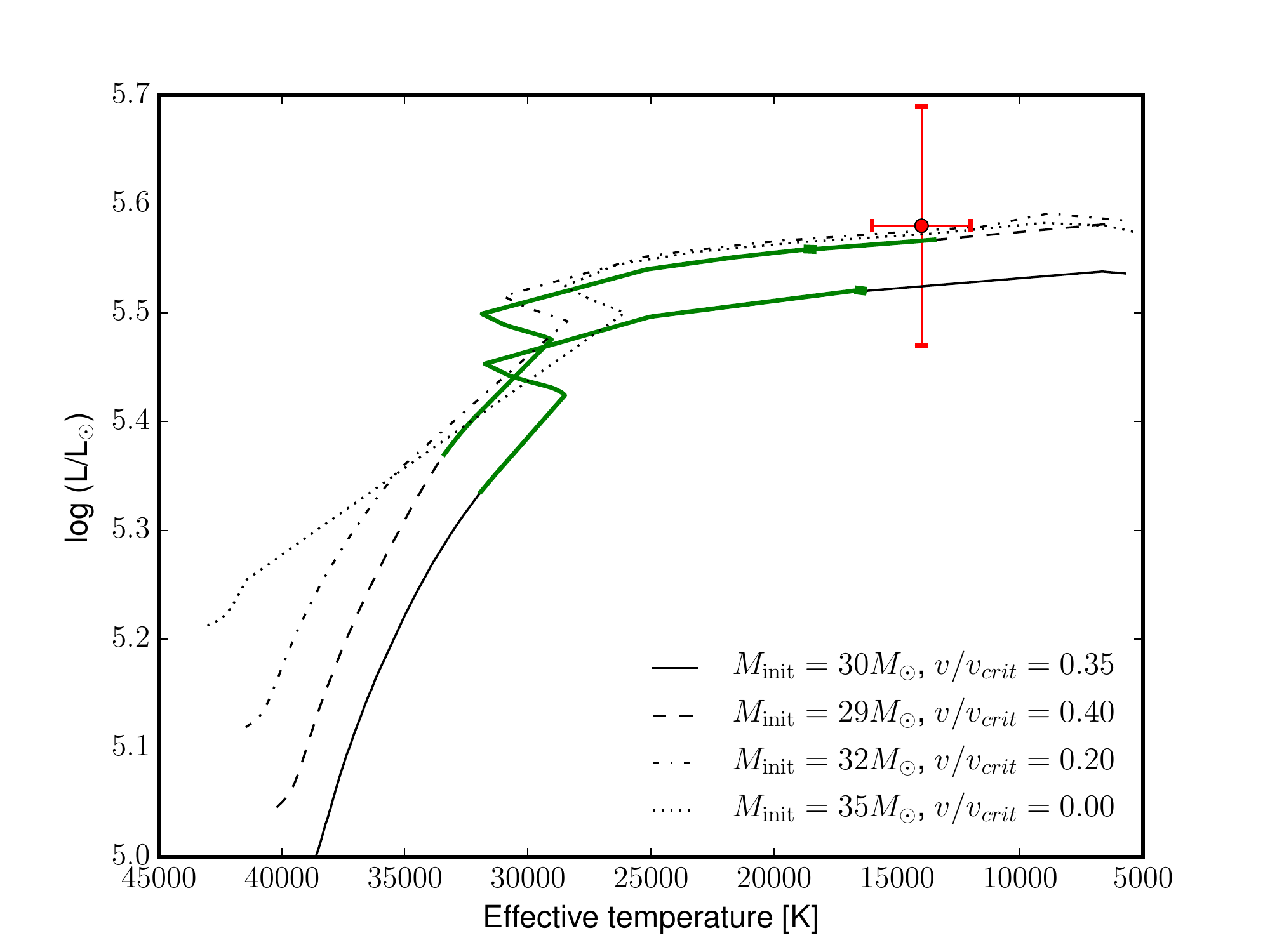}
\caption{Location of HD\,168625 (red) in the HR diagram and location at the time of the nebula ejection estimated from the abundances of the nebula (green). The bold green lines specify the location where, in addition, the mass-loss rates is between $\log \dot{M} = -5.0$ and $\log \dot{M} = -4.6$. Evolutionary tracks are from \citet{ekstrom12}.}\label{fig:HR}
\end{figure}

In the present analysis, no clear SB2 signature was found in the optical spectrum of HD\,168625 but only variations of its radial velocities. We thus considered that the spectrum only comes from one star and that these variations come from pulsations \citep[see, Sher\,25,][]{taylor14}. We determined its \lL, \teff\ and \logg\ to be $5.58\pm 0.11$, $14000 \pm 2000$~K and $1.74 \pm 0.05$, respectively. From these parameters, we can estimate the location of the star on the Hertzsprung-Russell (HR) diagram and associate this location with an evolutionary track from \citet{ekstrom12}. We also determine from the abundances that we derive the location where the nebula should have been ejected. We display in Fig.~\ref{fig:HR} the current location of the star and the "predicted" location of the star at the time of the ejection of the nebula. Both locations have been compared to several evolutionary tracks computed for different masses and different rotation rates. It appears that the observations agree, within the uncertainties, with a star of initial masses between 28 and 33\msun\ and a critical velocity $v/v_{crit}$ between 0.3 and 0.4. The tracks computed without rotation or with a small rotational rate ($v/v_{crit}<0.3$) fail at reproducing the abundances found in the nebula and in the star (no green line is observed in Fig~\ref{fig:HR} for those tracks).

\begin{figure}
\centering 
\includegraphics[width=8cm,bb=30 5 541 403,clip]{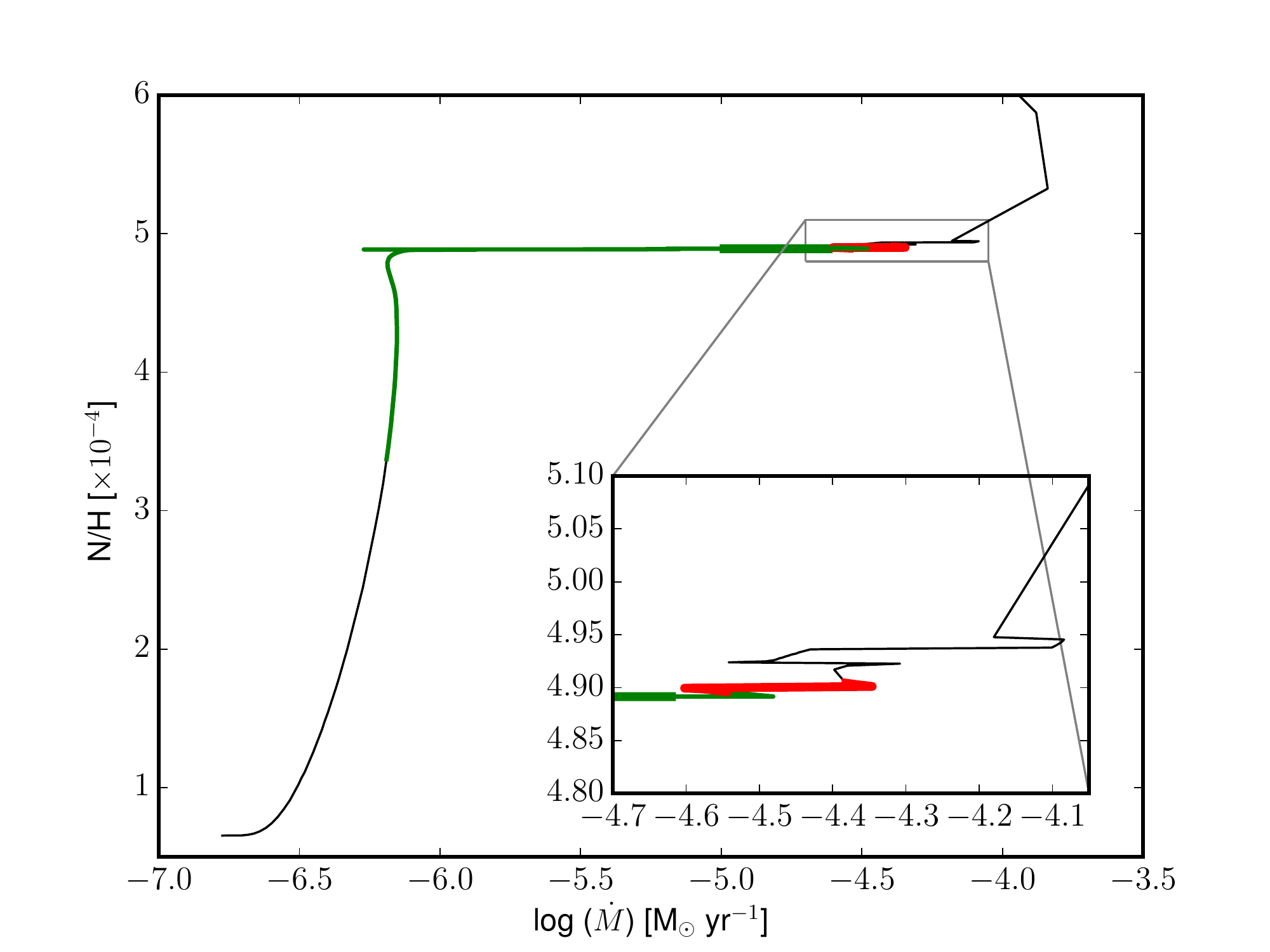}
\caption{Evolution of the N abundance (in number) as a function of mass-loss rate for a $30\,\msun$ star of solar metallicity and an initial rotational rate of 0.35. The evolutionary track comes from \citet{ekstrom12}. The green line corresponds to the nitrogen abundance determined for the nebula from the {\it Herschel}/PACS data and the bold green line specifies where $\log \dot{M}$ is between $-5.0$ and $-4.6$. The red line represents the location of the star according to the stellar N content and the stellar parameters from the CMFGEN model.}\label{fig:N_Mdot}
\end{figure}

\begin{figure}
\centering 
\includegraphics[width=8cm,bb=25 5 541 403,clip]{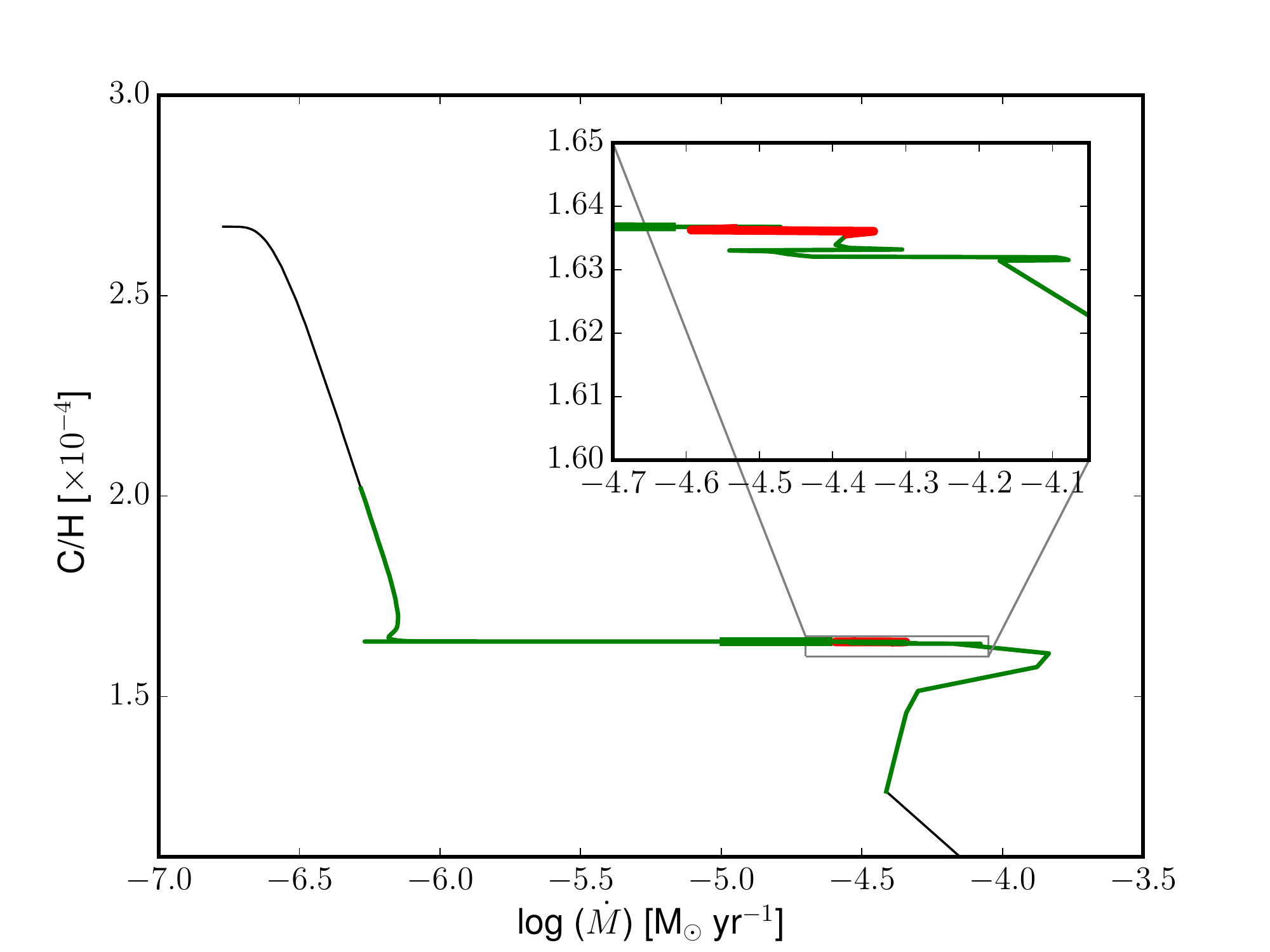}
\caption{Same as for Fig.~\ref{fig:N_Mdot} but for carbon abundance.}\label{fig:C_Mdot}
\end{figure}

The N/H and the C/H abundances versus the theoretical mass-loss rate for a star with an initial mass $M_{\mathrm{init}} = 30\,\msun$ and a $v/v_{crit} = 0.35$ are exhibited in Fig.~\ref{fig:N_Mdot} and in Fig.~\ref{fig:C_Mdot}, respectively. To create those figures, we take into account the values of the N/H and the C/H abundances that we derive from our analysis but also the stellar parameters \lL\ and \teff\ obtained from the best fit CMFGEN model. The size of the nebular shell (from the largest shell, 1.15\,pc, to the ionized nebula, 0.22\,pc), its total mass ($M_i + M_H + M_{\mathrm{dust}} \sim 0.17\,\msun + 0.99\,\msun \sim 1.17 \pm 0.34\,\msun$, the mass of the dust shell being negligible) and its expansion velocity (19~\kms, \citealt{pasquali02}) give a mass-loss rate between $\log \dot{M} = -5.0$ and $\log \dot{M} = -4.6$. This allows us to refine the epoch of the ejection of the nebula (bold green line in Figs~\ref{fig:HR},~\ref{fig:N_Mdot}, and~\ref{fig:C_Mdot}). This epoch corresponds to the end of the main-sequence phase, and, from the current location of the central star, it also appears that the star has not yet reached the red supergiant (RSG) phase. We note however that possible rapid excursions toward higher mass-loss rates that are not expected by current evolutionary tracks could also explain such values.

\subsection{Evolutionary scenarios}
\subsubsection{Single star scenario}

Single star evolutionary tracks can explain the evolution of the N content between the nebula and the central star but this still does not explain the shape and the different shells observed of the nebula around HD\,168625. A few objects exhibit an analogue bipolar structure to SN\,1987A \citep{plait95,mattila10}. Similar polar empty rings have been observed for Sher\,25 \citep{hendry08}, SWB1 \citep{smith13}, or MN18 \citep{gvaramadze15} but the nature of their formation still remains unknown.

It is however worth noticing that none of these objects has reached the RSG phase. \citet{smartt02} indeed showed from the determination of the surface abundances of Sher\,25 that this star has not gone through a RSG phase. \citet{smith13} estimated the N content of the ring of SWB1 to the solar value and came to the conclusion that this star has also not gone through a RSG phase. \citet{gvaramadze15}, from a CMFGEN analysis, determined that MN18 is a redward evolving star that became a BSG only recently. Therefore, the scenario proposed by \citet{chita08} aiming at explaining the creation of the bipolar structure from interactions between a BSG wind and a RSG thin shell seems thus compromised. Moreover, this theory fails at representing the empty polar rings.

From our analysis, nothing prevents us to think that the larger polar rings (detected on the 8\,\mum\ image) have the same age as the equatorial ring/torus. In this case, as already mentioned by \citet{smith07}, they could not have been ejected in a RSG phase because they would be far too fast. Therefore, since the creation of these two structures (the equatorial ring and the polar caps) suggests a common ejection occuring during the BSG phase following the main sequence.  According to \citet{smith13}, the polar rings would be formed by a dusty ionized flow of gas photoevaporated from the equatorial dense ring/torus. Its pressure prevents the stellar wind from moving towards the equatorial shell so that it would be deviated toward the poles, creating the polar empty rings. 

Rotation has also been claimed as a possible cause to explain the formation of the bipolar structure \citep[see e.g.,][]{gvaramadze15}. This scenario assumes that a fast stellar wind ($\vinf \sim 350~\kms$ for HD\,168625) concentrated in the equatorial plane because of the rotation collides with a slower and denser wind that moves away from the star at smaller velocities (i.e., 19~\kms for HD\,168625). The outflow is then collimated in the polar direction by the presence of this dense equatorial material. The disk is however expected by assuming a large rotation velocity, which is not the case for HD\,168625 ($\vsini = 60\,\kms$) \citep[see][for further details]{bjorkman93}. Furthermore, hydrodynamical simulations were performed by \citet[][and the references therein]{smith13} and failed at representing empty polar rings, but rather complete polar caps.

Finally, it is also possible that the illumination of the different parts of the nebula is anisotropic, indicating that some parts of the nebula could block some lines-of-sight, and would expect to result in a bipolar appearance.

\subsubsection{Binary system scenario}

Although no clear evidence has been found that HD\,168625 is a binary system, the detection of the presumably wide-orbit companion by \citet{pasquali02} and \citet{martayan16} (even though no clear motion of this "companion" is detected between the two epochs) and the possible role of a binary/multiple system in the evolution of this object remain questionable. Moreover, HD\,168625 is quoted as a progenitor of SN\,1987A that has been thought to be a binary merger by \citet{morris09}. Therefore, we must envisage the binarity scenario to explain the ejection of this bipolar structure. The binary evolutionary tracks being complex to calculate, we were not able to compare the CNO abundances that we have determined for the central star and for the nebula to those tracks to see whether a match could be found. 

As the abundances determined for the central star are in agreement with those estimated from a single star evolutionary track, it does not allow us to consider a mass transfer by Roche lobe overflow. Even though such a transfer happened, we expect that the rotational velocity of the mass gainer increases whilst that of the mass donor decreases. Given the projected rotational velocity, it seems unlikely that HD\,168265 is the gainer unless it loses a significant quantity of angular momentum on a very short time-scale. We indeed see that, for HD\,168625, the projected rotational velocity that we have determined from the optical spectrum is not large enough to classify it as rapid rotator. If we assume an inclination of about $40\degr$ or $60\degr$ (both values estimated from the imaging analysis), the rotational velocity of HD\,168625 is about 70--95~\kms. We also see that, for the other possible SN progenitors that show similar nebulae in the polar direction, their projected rotational velocity is also quite small. As outlined by \citet{gvaramadze15}, the most efficient mechanisms to spin down a massive star is the magnetic braking via a magnetized wind but that mechanism requires magnetic fields of about $10^3 - 10^4$~G and a high mass-loss rate. HD\,168625 has been probed by \citet{fossati15} to look for magnetic field but these authors did not detect any significant magnetic field. If a transfer of material by Roche lobe overflow occurs in HD\,168625, we should also see emissions in the X-ray domain. However, \citet{naze12} did not detect any significant X-ray emission for this star. Therefore, we must consider another form of exchange: the binary merger. In this case, this scenario could explain the absence of X-ray but would imply rapid rotation, which is not found in HD\,168625.


\section{Conclusion}
\label{sec:conc}

We have presented the analysis of the {\it Herschel}/PACS imaging and spectroscopic data of the nebula surrounding the candidate LBV HD\,168625 together with H$\alpha$ and mid-infrared images as well as high-resolution optical spectra. The images, covering the H$\alpha$--160\,\mum\ domain, confirm the complex morphology of the nebula, composed of torus-like shells visible at different wavelengths and of a perpendicular bipolar structure that is observed in the H$\alpha$, 8\,\mum\ and 13\,\mum\ images. 

We have also seen that the ionized nebula is smaller than the entire nebular shell. We determined in the present paper that about 85\% of the total mass of the nebula comes from the neutral gas ($M_H \sim  1.0\,\msun$ out of 1.17\,\msun\ for the total mass) against 14\% from the ionized shell ($M_i \sim 0.17\,\msun$), the remaining percentage being attributed to the dust. 

The far-infrared spectrum of the nebula provided us with the nitrogen and carbon abundances. We have determined an overabundance in nitrogen (N/H~$\cong 4.1 \times~10^{-4}$) and a depletion in carbon (C/H~$\cong 1.6 \times~10^{-4}$) in the nebula around HD\,168625. We compared these values to the CNO abundances determined with the CMFGEN atmosphere models computed from optical spectra of the central star. We were able to constrain the evolutionary path of the star and to determine the moment of the nebula ejection. This star seems to have an initial mass between $28$ and $33\,\msun$ and should have ejected its material during or just after the Blue Supergiant phase and before the red supergiant stage. The evolution of HD\,168625 seems thus related to the evolution of a single star even though the complex structure of its bipolar/biconal rings remains unexplained under this assumption and could therefore favor the binary scenario.


\begin{acknowledgements}
We thank the anonymous referee for his/her comments and remarks. We also thank Prof. Gregor Rauw for their valuable comments on the  manuscript and Dr. Nick Cox for providing the scanamorphos images. L.M., D.H., P.R. acknowledge support from the Belgian Federal Science Policy Office via the PRODEX Program of ESA. The Li{\`e}ge team also acknowledges support from the Fonds National de la Recherche Scientifique (F.R.S.-F.N.R.S.). This research was also funded through the ARC grant for Concerted Research Actions, financed by the French Community of Belgium (Wallonia-Brussels Federation). PACS has been developed by a consortium of institutes led by MPE (Germany) and including UVIE (Austria); KU Leuven, CSL, IMEC (Belgium); CEA, LAM (France); MPIA (Germany); INAF-IFSI/OAA/OAP/OAT, LENS, SISSA(Italy); IAC (Spain). This development has been supported by the funding agencies BMVIT (Austria), ESA-PRODEX (Belgium), CEA/CNES (France), DLR(Germany), ASI/INAF (Italy), and CICYT/MCYT (Spain). Data presented in this paper were analyzed using "HIPE", a joint development by the {\it Herschel} Science Ground Segment Consortium, consisting of ESA, the NASA {\it Herschel} Science Center, and the HIFI, PACS and SPIRE consortia. This research has made use of the NASA/IPAC Infrared Science Archive, which is operated by the Jet Propulsion Laboratory, California Institute of Technology, as well as NASA/ADS and SIMBAD (CDS/Strasbourg) databases.
\end{acknowledgements}

   \bibliography{LBV}

\begin{thebibliography}{68}
\expandafter\ifx\csname natexlab\endcsname\relax\def\natexlab#1{#1}\fi

\bibitem[{{Aldoretta} {et~al.}(2015){Aldoretta}, {Caballero-Nieves}, {Gies},
  {Nelan}, {Wallace}, {Hartkopf}, {Henry}, {Jao}, {Ma{\'{\i}}z Apell{\'a}niz},
  {Mason}, {Moffat}, {Norris}, {Richardson}, \& {Williams}}]{aldoretta15}
{Aldoretta}, E.~J., {Caballero-Nieves}, S.~M., {Gies}, D.~R., {et~al.} 2015,
  \aj, 149, 26

\bibitem[{{Bjorkman} \& {Cassinelli}(1993)}]{bjorkman93}
{Bjorkman}, J.~E. \& {Cassinelli}, J.~P. 1993, \apj, 409, 429

\bibitem[{{Blommaert} {et~al.}(2014){Blommaert}, {de Vries}, {Waters},
  {Waelkens}, {Min}, {Van Winckel}, {Molster}, {Decin}, {Groenewegen},
  {Barlow}, {Garc{\'{\i}}a-Lario}, {Kerschbaum}, {Posch}, {Royer}, {Ueta},
  {Vandenbussche}, {Van de Steene}, \& {van Hoof}}]{blommaert14}
{Blommaert}, J.~A.~D.~L., {de Vries}, B.~L., {Waters}, L.~B.~F.~M., {et~al.}
  2014, \aap, 565, A109

\bibitem[{{Brand} \& {Blitz}(1993)}]{brand93}
{Brand}, J. \& {Blitz}, L. 1993, \aap, 275, 67

\bibitem[{{Chentsov} \& {Luud}(1989)}]{chentsov89}
{Chentsov}, E.~L. \& {Luud}, L. 1989, Astrophysics, 31, 415

\bibitem[{{Chita} {et~al.}(2008){Chita}, {Langer}, {van Marle},
  {Garc{\'{\i}}a-Segura}, \& {Heger}}]{chita08}
{Chita}, S.~M., {Langer}, N., {van Marle}, A.~J., {Garc{\'{\i}}a-Segura}, G.,
  \& {Heger}, A. 2008, \aap, 488, L37

\bibitem[{{Damineli} {et~al.}(1997){Damineli}, {Conti}, \&
  {Lopes}}]{daminelli97}
{Damineli}, A., {Conti}, P.~S., \& {Lopes}, D.~F. 1997, \na, 2, 107

\bibitem[{{Draine}(2011)}]{draine11}
{Draine}, B.~T. 2011, {Physics of the Interstellar and Intergalactic Medium}

\bibitem[{{Ducati}(2002)}]{ducati02}
{Ducati}, J.~R. 2002, VizieR Online Data Catalog, 2237, 0

\bibitem[{{Ekstr{\"o}m} {et~al.}(2012){Ekstr{\"o}m}, {Georgy}, {Eggenberger},
  {Meynet}, {Mowlavi}, {Wyttenbach}, {Granada}, {Decressin}, {Hirschi},
  {Frischknecht}, {Charbonnel}, \& {Maeder}}]{ekstrom12}
{Ekstr{\"o}m}, S., {Georgy}, C., {Eggenberger}, P., {et~al.} 2012, \aap, 537,
  A146

\bibitem[{{Fazio} {et~al.}(2004){Fazio}, {Hora}, {Allen}, {Ashby}, {Barmby},
  {Deutsch}, {Huang}, {Kleiner}, {Marengo}, {Megeath}, {Melnick}, {Pahre},
  {Patten}, {Polizotti}, {Smith}, {Taylor}, {Wang}, {Willner}, {Hoffmann},
  {Pipher}, {Forrest}, {McMurty}, {McCreight}, {McKelvey}, {McMurray}, {Koch},
  {Moseley}, {Arendt}, {Mentzell}, {Marx}, {Losch}, {Mayman}, {Eichhorn},
  {Krebs}, {Jhabvala}, {Gezari}, {Fixsen}, {Flores}, {Shakoorzadeh}, {Jungo},
  {Hakun}, {Workman}, {Karpati}, {Kichak}, {Whitley}, {Mann}, {Tollestrup},
  {Eisenhardt}, {Stern}, {Gorjian}, {Bhattacharya}, {Carey}, {Nelson},
  {Glaccum}, {Lacy}, {Lowrance}, {Laine}, {Reach}, {Stauffer}, {Surace},
  {Wilson}, {Wright}, {Hoffman}, {Domingo}, \& {Cohen}}]{fazio04}
{Fazio}, G.~G., {Hora}, J.~L., {Allen}, L.~E., {et~al.} 2004, \apjs, 154, 10

\bibitem[{{Fossati} {et~al.}(2015){Fossati}, {Castro}, {Sch{\"o}ller},
  {Hubrig}, {Langer}, {Morel}, {Briquet}, {Herrero}, {Przybilla}, {Sana},
  {Schneider}, {de Koter}, \& {BOB Collaboration}}]{fossati15}
{Fossati}, L., {Castro}, N., {Sch{\"o}ller}, M., {et~al.} 2015, \aap, 582, A45

\bibitem[{{Garc{\'{\i}}a-Lario} {et~al.}(2001){Garc{\'{\i}}a-Lario},
  {Sivarani}, {Parthasarathy}, \& {Manchado}}]{garcia-lario01}
{Garc{\'{\i}}a-Lario}, P., {Sivarani}, T., {Parthasarathy}, M., \& {Manchado},
  A. 2001, in Astrophysics and Space Science Library, Vol. 265, Astrophysics
  and Space Science Library, ed. R.~{Szczerba} \& S.~K. {G{\'o}rny}

\bibitem[{{Grevesse} {et~al.}(2010){Grevesse}, {Asplund}, {Sauval}, \&
  {Scott}}]{grevesse10}
{Grevesse}, N., {Asplund}, M., {Sauval}, A.~J., \& {Scott}, P. 2010, \apss,
  328, 179

\bibitem[{{Groenewegen} {et~al.}(2011){Groenewegen}, {Waelkens}, {Barlow},
  {Kerschbaum}, {Garcia-Lario}, {Cernicharo}, {Blommaert}, {Bouwman}, {Cohen},
  {Cox}, {Decin}, {Exter}, {Gear}, {Gomez}, {Hargrave}, {Henning},
  {Hutsem{\'e}kers}, {Ivison}, {Jorissen}, {Krause}, {Ladjal}, {Leeks}, {Lim},
  {Matsuura}, {Naz{\'e}}, {Olofsson}, {Ottensamer}, {Polehampton}, {Posch},
  {Rauw}, {Royer}, {Sibthorpe}, {Swinyard}, {Ueta}, {Vamvatira-Nakou},
  {Vandenbussche}, {van de Steene}, {van Eck}, {van Hoof}, {van Winckel},
  {Verdugo}, \& {Wesson}}]{groenewegen11}
{Groenewegen}, M.~A.~T., {Waelkens}, C., {Barlow}, M.~J., {et~al.} 2011, \aap,
  526, A162

\bibitem[{{Gvaramadze} {et~al.}(2015){Gvaramadze}, {Kniazev}, {Bestenlehner},
  {Bodensteiner}, {Langer}, {Greiner}, {Grebel}, {Berdnikov}, \&
  {Beletsky}}]{gvaramadze15}
{Gvaramadze}, V.~V., {Kniazev}, A.~Y., {Bestenlehner}, J.~M., {et~al.} 2015,
  \mnras, 454, 219

\bibitem[{{Hendry} {et~al.}(2008){Hendry}, {Smartt}, {Skillman}, {Evans},
  {Trundle}, {Lennon}, {Crowther}, \& {Hunter}}]{hendry08}
{Hendry}, M.~A., {Smartt}, S.~J., {Skillman}, E.~D., {et~al.} 2008, \mnras,
  388, 1127

\bibitem[{{Hillier} \& {Miller}(1998)}]{hillier98}
{Hillier}, D.~J. \& {Miller}, D.~L. 1998, \apj, 496, 407

\bibitem[{{Hollenbach} \& {McKee}(1989)}]{hollenback89}
{Hollenbach}, D. \& {McKee}, C.~F. 1989, \apj, 342, 306

\bibitem[{{Humphreys} \& {Davidson}(1994)}]{humphreys94}
{Humphreys}, R.~M. \& {Davidson}, K. 1994, \pasp, 106, 1025

\bibitem[{{Hutsem{\'e}kers} {et~al.}(1994){Hutsem{\'e}kers}, {van Drom},
  {Gosset}, \& {Melnick}}]{hutsemekers94}
{Hutsem{\'e}kers}, D., {van Drom}, E., {Gosset}, E., \& {Melnick}, J. 1994,
  \aap, 290, 906

\bibitem[{{Jim{\'e}nez-Esteban} {et~al.}(2010){Jim{\'e}nez-Esteban}, {Rizzo},
  \& {Palau}}]{jimenez10}
{Jim{\'e}nez-Esteban}, F.~M., {Rizzo}, J.~R., \& {Palau}, A. 2010, \apj, 713,
  429

\bibitem[{{Kaufman} {et~al.}(1999){Kaufman}, {Wolfire}, {Hollenbach}, \&
  {Luhman}}]{kaufman99}
{Kaufman}, M.~J., {Wolfire}, M.~G., {Hollenbach}, D.~J., \& {Luhman}, M.~L.
  1999, \apj, 527, 795

\bibitem[{{Leitherer} {et~al.}(1995){Leitherer}, {Chapman}, \&
  {Koribalski}}]{leitherer95}
{Leitherer}, C., {Chapman}, J.~M., \& {Koribalski}, B. 1995, \apj, 450, 289

\bibitem[{{Lobel} {et~al.}(2013){Lobel}, {Groh}, {Martayan}, {Fr{\'e}mat},
  {Torres Dozinel}, {Raskin}, {Van Winckel}, {Prins}, {Pessemier}, {Waelkens},
  {Hensberge}, {Dumortier}, {Jorissen}, {Van Eck}, \& {Lehmann}}]{lobel13}
{Lobel}, A., {Groh}, J.~H., {Martayan}, C., {et~al.} 2013, \aap, 559, A16

\bibitem[{{Malhotra} {et~al.}(2001){Malhotra}, {Kaufman}, {Hollenbach},
  {Helou}, {Rubin}, {Brauher}, {Dale}, {Lu}, {Lord}, {Stacey}, {Contursi},
  {Hunter}, \& {Dinerstein}}]{malhorta01}
{Malhotra}, S., {Kaufman}, M.~J., {Hollenbach}, D., {et~al.} 2001, \apj, 561,
  766

\bibitem[{{Martayan} {et~al.}(2016){Martayan}, {Lobel}, {Baade}, {Mehner},
  {Rivinius}, {Boffin}, {Girard}, {Mawet}, {Montagnier}, {Blomme}, {Kervella},
  {Sana}, {{\v S}tefl}, {Zorec}, {Lacour}, {Le Bouquin}, {Martins},
  {M{\'e}rand}, {Patru}, {Selman}, \& {Fr{\'e}mat}}]{martayan16}
{Martayan}, C., {Lobel}, A., {Baade}, D., {et~al.} 2016, ArXiv e-prints
  [\eprint[arXiv]{1601.03542}]

\bibitem[{{Martins}(2011)}]{martins11}
{Martins}, F. 2011, Bulletin de la Societe Royale des Sciences de Liege, 80, 29

\bibitem[{{Martins} {et~al.}(2015){Martins}, {Herv{\'e}}, {Bouret},
  {Marcolino}, {Wade}, {Neiner}, {Alecian}, {Grunhut}, \& {Petit}}]{martins15}
{Martins}, F., {Herv{\'e}}, A., {Bouret}, J.-C., {et~al.} 2015, \aap, 575, A34

\bibitem[{{Martins} \& {Plez}(2006)}]{martins06}
{Martins}, F. \& {Plez}, B. 2006, \aap, 457, 637

\bibitem[{{Mattila} {et~al.}(2010){Mattila}, {Lundqvist}, {Gr{\"o}ningsson},
  {Meikle}, {Stathakis}, {Fransson}, \& {Cannon}}]{mattila10}
{Mattila}, S., {Lundqvist}, P., {Gr{\"o}ningsson}, P., {et~al.} 2010, \apj,
  717, 1140

\bibitem[{{Morgan} {et~al.}(1955){Morgan}, {Code}, \& {Whitford}}]{morgan55}
{Morgan}, W.~W., {Code}, A.~D., \& {Whitford}, A.~E. 1955, \apjs, 2, 41

\bibitem[{{Morris} \& {Podsiadlowski}(2009)}]{morris09}
{Morris}, T. \& {Podsiadlowski}, P. 2009, \mnras, 399, 515

\bibitem[{{Naz{\'e}} {et~al.}(2012){Naz{\'e}}, {Rauw}, \&
  {Hutsem{\'e}kers}}]{naze12}
{Naz{\'e}}, Y., {Rauw}, G., \& {Hutsem{\'e}kers}, D. 2012, \aap, 538, A47

\bibitem[{{Nota} {et~al.}(1996){Nota}, {Pasquali}, {Clampin}, {Pollacco},
  {Scuderi}, \& {Livio}}]{nota96}
{Nota}, A., {Pasquali}, A., {Clampin}, M., {et~al.} 1996, \apj, 473, 946

\bibitem[{{O'Hara} {et~al.}(2003){O'Hara}, {Meixner}, {Speck}, {Ueta}, \&
  {Bobrowsky}}]{ohara03}
{O'Hara}, T.~B., {Meixner}, M., {Speck}, A.~K., {Ueta}, T., \& {Bobrowsky}, M.
  2003, \apj, 598, 1255

\bibitem[{{Ott}(2010)}]{ott10}
{Ott}, S. 2010, in Astronomical Society of the Pacific Conference Series, Vol.
  434, Astronomical Data Analysis Software and Systems XIX, ed. Y.~{Mizumoto},
  K.-I. {Morita}, \& M.~{Ohishi}, 139

\bibitem[{{Pasquali} {et~al.}(2002){Pasquali}, {Nota}, {Smith}, {Akiyama},
  {Messineo}, \& {Clampin}}]{pasquali02}
{Pasquali}, A., {Nota}, A., {Smith}, L.~J., {et~al.} 2002, \aj, 124, 1625

\bibitem[{{Pilbratt} {et~al.}(2010){Pilbratt}, {Riedinger}, {Passvogel},
  {Crone}, {Doyle}, {Gageur}, {Heras}, {Jewell}, {Metcalfe}, {Ott}, \&
  {Schmidt}}]{pilbratt10}
{Pilbratt}, G.~L., {Riedinger}, J.~R., {Passvogel}, T., {et~al.} 2010, \aap,
  518, L1

\bibitem[{{Plait} {et~al.}(1995){Plait}, {Lundqvist}, {Chevalier}, \&
  {Kirshner}}]{plait95}
{Plait}, P.~C., {Lundqvist}, P., {Chevalier}, R.~A., \& {Kirshner}, R.~P. 1995,
  \apj, 439, 730

\bibitem[{{Poglitsch} {et~al.}(2010){Poglitsch}, {Waelkens}, {Geis},
  {Feuchtgruber}, {Vandenbussche}, {Rodriguez}, {Krause}, {Renotte}, {van
  Hoof}, {Saraceno}, {Cepa}, {Kerschbaum}, {Agn{\`e}se}, {Ali}, {Altieri},
  {Andreani}, {Augueres}, {Balog}, {Barl}, {Bauer}, {Belbachir}, {Benedettini},
  {Billot}, {Boulade}, {Bischof}, {Blommaert}, {Callut}, {Cara}, {Cerulli},
  {Cesarsky}, {Contursi}, {Creten}, {De Meester}, {Doublier}, {Doumayrou},
  {Duband}, {Exter}, {Genzel}, {Gillis}, {Gr{\"o}zinger}, {Henning},
  {Herreros}, {Huygen}, {Inguscio}, {Jakob}, {Jamar}, {Jean}, {de Jong},
  {Katterloher}, {Kiss}, {Klaas}, {Lemke}, {Lutz}, {Madden}, {Marquet},
  {Martignac}, {Mazy}, {Merken}, {Montfort}, {Morbidelli}, {M{\"u}ller},
  {Nielbock}, {Okumura}, {Orfei}, {Ottensamer}, {Pezzuto}, {Popesso},
  {Putzeys}, {Regibo}, {Reveret}, {Royer}, {Sauvage}, {Schreiber}, {Stegmaier},
  {Schmitt}, {Schubert}, {Sturm}, {Thiel}, {Tofani}, {Vavrek}, {Wetzstein},
  {Wieprecht}, \& {Wiezorrek}}]{poglitsch10}
{Poglitsch}, A., {Waelkens}, C., {Geis}, N., {et~al.} 2010, \aap, 518, L2

\bibitem[{{Popper} \& {Seyfert}(1940)}]{popper40}
{Popper}, D.~M. \& {Seyfert}, C.~K. 1940, \pasp, 52, 401

\bibitem[{{Rieke} {et~al.}(2004){Rieke}, {Young}, {Engelbracht}, {Kelly},
  {Low}, {Haller}, {Beeman}, {Gordon}, {Stansberry}, {Misselt}, {Cadien},
  {Morrison}, {Rivlis}, {Latter}, {Noriega-Crespo}, {Padgett}, {Stapelfeldt},
  {Hines}, {Egami}, {Muzerolle}, {Alonso-Herrero}, {Blaylock}, {Dole}, {Hinz},
  {Le Floc'h}, {Papovich}, {P{\'e}rez-Gonz{\'a}lez}, {Smith}, {Su}, {Bennett},
  {Frayer}, {Henderson}, {Lu}, {Masci}, {Pesenson}, {Rebull}, {Rho}, {Keene},
  {Stolovy}, {Wachter}, {Wheaton}, {Werner}, \& {Richards}}]{rieke04}
{Rieke}, G.~H., {Young}, E.~T., {Engelbracht}, C.~W., {et~al.} 2004, \apjs,
  154, 25

\bibitem[{{Rivinius} {et~al.}(2015){Rivinius}, {Boffin}, {de Wit}, {Mehner},
  {Martayan}, {Guieu}, \& {Le Bouquin}}]{rivinius15}
{Rivinius}, T., {Boffin}, H.~M.~J., {de Wit}, W.~J., {et~al.} 2015, in IAU
  Symposium, Vol. 307, IAU Symposium, ed. G.~{Meynet}, C.~{Georgy}, J.~{Groh},
  \& P.~{Stee}, 295--296

\bibitem[{{Robberto} \& {Herbst}(1998)}]{robberto98}
{Robberto}, M. \& {Herbst}, T.~M. 1998, \apj, 498, 400

\bibitem[{{Roussel}(2013)}]{roussel13}
{Roussel}, H. 2013, \pasp, 125, 1126

\bibitem[{{Rubin} {et~al.}(1994){Rubin}, {Simpson}, {Lord}, {Colgan},
  {Erickson}, \& {Haas}}]{rubin94}
{Rubin}, R.~H., {Simpson}, J.~P., {Lord}, S.~D., {et~al.} 1994, \apj, 420, 772

\bibitem[{{Sana} {et~al.}(2006){Sana}, {Gosset}, \& {Rauw}}]{sana06}
{Sana}, H., {Gosset}, E., \& {Rauw}, G. 2006, \mnras, 371, 67

\bibitem[{{Schmidt-Kaler}(1982)}]{schmidtkaler82}
{Schmidt-Kaler}, T. 1982, in: Landolt-Börnstein, Numerical Data and Functional
  Relationships in Science and Technology, Vol.~2, Stars and Star Clusters /
  Sterne und Sternhaufen, ed. K.~{Schaifers} \& H.~H. {Voigt} (Berlin:
  Springer-Verlag), 451

\bibitem[{{Sim{\'o}n-D{\'{\i}}az} \& {Herrero}(2007)}]{simondiaz07}
{Sim{\'o}n-D{\'{\i}}az}, S. \& {Herrero}, A. 2007, \aap, 468, 1063

\bibitem[{{Skinner}(1997)}]{skinner97}
{Skinner}, C.~J. 1997, in Astronomical Society of the Pacific Conference
  Series, Vol. 120, Luminous Blue Variables: Massive Stars in Transition, ed.
  A.~{Nota} \& H.~{Lamers}, 322

\bibitem[{{Smartt} {et~al.}(2002){Smartt}, {Lennon}, {Kudritzki}, {Rosales},
  {Ryans}, \& {Wright}}]{smartt02}
{Smartt}, S.~J., {Lennon}, D.~J., {Kudritzki}, R.~P., {et~al.} 2002, \aap, 391,
  979

\bibitem[{{Smith}(2007)}]{smith07}
{Smith}, N. 2007, \aj, 133, 1034

\bibitem[{{Smith} {et~al.}(2013){Smith}, {Arnett}, {Bally}, {Ginsburg}, \&
  {Filippenko}}]{smith13}
{Smith}, N., {Arnett}, W.~D., {Bally}, J., {Ginsburg}, A., \& {Filippenko},
  A.~V. 2013, \mnras, 429, 1324

\bibitem[{{Sterken} {et~al.}(1999){Sterken}, {Arentoft}, {Duerbeck}, \&
  {Brogt}}]{sterken99}
{Sterken}, C., {Arentoft}, T., {Duerbeck}, H.~W., \& {Brogt}, E. 1999, \aap,
  349, 532

\bibitem[{{Taylor} {et~al.}(2014){Taylor}, {Evans}, {Sim{\'o}n-D{\'{\i}}az},
  {Sana}, {Langer}, {Smith}, \& {Smartt}}]{taylor14}
{Taylor}, W.~D., {Evans}, C.~J., {Sim{\'o}n-D{\'{\i}}az}, S., {et~al.} 2014,
  \mnras, 442, 1483

\bibitem[{{Tielens}(2005)}]{tielen05}
{Tielens}, A.~G.~G.~M. 2005, {The Physics and Chemistry of the Interstellar
  Medium}

\bibitem[{{Tielens} \& {Hollenbach}(1985)}]{tielens85}
{Tielens}, A.~G.~G.~M. \& {Hollenbach}, D. 1985, \apj, 291, 722

\bibitem[{{Umana} {et~al.}(2010){Umana}, {Buemi}, {Trigilio}, {Leto}, \&
  {Hora}}]{umana10}
{Umana}, G., {Buemi}, C.~S., {Trigilio}, C., {Leto}, P., \& {Hora}, J.~L. 2010,
  \apj, 718, 1036

\bibitem[{{Vamvatira-Nakou} {et~al.}(2015){Vamvatira-Nakou}, {Hutsemekers},
  {Royer}, {Cox}, {Naze}, {Rauw}, {Waelkens}, \& {Groenewegen}}]{vamvatira15}
{Vamvatira-Nakou}, C., {Hutsemekers}, D., {Royer}, P., {et~al.} 2015, ArXiv
  e-prints [\eprint[arXiv]{1504.03204}]

\bibitem[{{Vamvatira-Nakou} {et~al.}(2013){Vamvatira-Nakou}, {Hutsem{\'e}kers},
  {Royer}, {Naz{\'e}}, {Magain}, {Exter}, {Waelkens}, \&
  {Groenewegen}}]{vamvatira13}
{Vamvatira-Nakou}, C., {Hutsem{\'e}kers}, D., {Royer}, P., {et~al.} 2013, \aap,
  557, A20

\bibitem[{{Vamvatira-Nakou} {et~al.}(2016){Vamvatira-Nakou}, {Hutsemekers},
  {Royer}, {Waelkens}, {Groenewegen}, \& {Barlow}}]{vamvatira16}
{Vamvatira-Nakou}, C., {Hutsemekers}, D., {Royer}, P., {et~al.} 2016, ArXiv
  e-prints [\eprint[arXiv]{1602.03422}]

\bibitem[{{van Genderen} {et~al.}(1992){van Genderen}, {van den Bosch},
  {Dessing}, {Fehmers}, {van Grunsven}, {van der Heiden}, {Janssens}, {Kalter},
  {van der Meer}, {van Ojik}, {Smit}, \& {Zijderveld}}]{vangenderen92}
{van Genderen}, A.~M., {van den Bosch}, F.~C., {Dessing}, F., {et~al.} 1992,
  \aap, 264, 88

\bibitem[{{van Marle} {et~al.}(2007){van Marle}, {Langer}, \&
  {Garc{\'{\i}}a-Segura}}]{vanmarle07}
{van Marle}, A.~J., {Langer}, N., \& {Garc{\'{\i}}a-Segura}, G. 2007, \aap,
  469, 941

\bibitem[{{Weis}(2011)}]{weis11}
{Weis}, K. 2011, in IAU Symposium, Vol. 272, Active OB Stars: Structure,
  Evolution, Mass Loss, and Critical Limits, ed. C.~{Neiner}, G.~{Wade},
  G.~{Meynet}, \& G.~{Peters}, 372--377

\bibitem[{{Weis}(2012)}]{weis12}
{Weis}, K. 2012, in Astrophysics and Space Science Library, Vol. 384, Eta
  Carinae and the Supernova Impostors, ed. K.~{Davidson} \& R.~M. {Humphreys},
  171

\bibitem[{{Wright} {et~al.}(2010){Wright}, {Eisenhardt}, {Mainzer}, {Ressler},
  {Cutri}, {Jarrett}, {Kirkpatrick}, {Padgett}, {McMillan}, {Skrutskie},
  {Stanford}, {Cohen}, {Walker}, {Mather}, {Leisawitz}, {Gautier}, {McLean},
  {Benford}, {Lonsdale}, {Blain}, {Mendez}, {Irace}, {Duval}, {Liu}, {Royer},
  {Heinrichsen}, {Howard}, {Shannon}, {Kendall}, {Walsh}, {Larsen}, {Cardon},
  {Schick}, {Schwalm}, {Abid}, {Fabinsky}, {Naes}, \& {Tsai}}]{wright10}
{Wright}, E.~L., {Eisenhardt}, P.~R.~M., {Mainzer}, A.~K., {et~al.} 2010, \aj,
  140, 1868

\bibitem[{{Young Owl} {et~al.}(2002){Young Owl}, {Meixner}, {Fong}, {Haas},
  {Rudolph}, \& {Tielens}}]{young02}
{Young Owl}, R.~C., {Meixner}, M.~M., {Fong}, D., {et~al.} 2002, \apj, 578, 885

\end{thebibliography}
\newpage
\begin{appendix}
\section{Journal of observation of the optical spectra}
\begin{table*}
\caption{Journal of observation of the optical spectra retrieved from the ESO archives.}             
\centering                          
\begin{tabular}{l c c c c}        
\hline\hline                 
Program ID & PI  & Instrument & Date [JJ MM YYYY]  & Exposure time [s]     \\    
\hline                        
069.D-0381(A) & Gosset & FEROS & 18 04 2002 & $1800$\\
069.D-0378(A) & Szeifert & FEROS & 29 06 2002 & $3\times~1800$\\
073.D-0609(A) & Sana   & FEROS & 05 05 2004  & $720$\\
&        &       &             & $480$\\
&        &       & 08 05 2004  & $480$\\
&        &       &             & $720$\\
079.B-0856(A) & Przybilla & FEROS & 14 07 2007 & $680$\\
082.D-0136(A) & Evans & FEROS & 19 03 2009  & $3 \times~240$\\
&        &       & 20 03 2009 & $5 \times~300$\\
&        &       & 21 03 2009 & $4 \times~600$\\
&        &       & 22 03 2009 & $2 \times~600$\\
&        &       & 23 03 2009 & $600$\\
&        &       & 24 03 2009 & $600$\\
266.D-5655(A) & Paranal Observatory & UVES & 10 07 2001 & \\
083.C-0503(A) & Iglesias-Groth & UVES & 15 07 2009 & \\
&                &      & 19 08 2009 & \\
093.D-0415(A) & Martayan & XShooter & 27 05 2014 &\\
& & & 02 06 2014 & \\
\hline                                   
\end{tabular}
\end{table*}

\section{Emission line fluxes for each spaxel}
\begin{table*}
\caption{Line fluxes in each spaxel. A dash indicates a poor signal-to-noise ratio or a non-detection.}             
\centering                          
\begin{tabular}{l |c |c |c |c |c}        
\hline\hline                 
Line & Flux  & Flux & Flux &  Flux &  Flux  \\
     & [$10^{-15}$ W m$^{-2}$] &[$10^{-15}$ W m$^{-2}$]&[$10^{-15}$ W m$^{-2}$]&[$10^{-15}$ W m$^{-2}$]&[$10^{-15}$ W m$^{-2}$]\\
\hline
&  spaxel (0,0) & spaxel (0,1) & spaxel (0,2) & spaxel (0,3) & spaxel (0,4)\\
\hline
$[\ion{O}{i}]\,63\,\mum$   & $0.01\pm0.01$  & $0.10 \pm 0.06$ & $0.10 \pm 0.05$  & $0.05 \pm 0.03$  & $0.02 \pm 0.02$ \\
$[\ion{N}{ii}]\,122\,\mum$ &  $0.01\pm0.01$ & $0.03 \pm 0.02$ & $0.05 \pm 0.03$  & $0.05 \pm 0.03$  & $0.02 \pm 0.02$ \\
$[\ion{O}{i}]\,146\,\mum$  &       --       &      --         &       --         & $0.02 \pm 0.01$  &        --       \\
$[\ion{C}{ii}]\,158\,\mum$ &  $0.13\pm0.07$ & $0.14 \pm 0.04$ & $0.14 \pm 0.06$  & $0.13 \pm 0.04$  & $0.09 \pm 0.04$ \\
$[\ion{N}{ii}]\,205\,\mum$ &       --       &      --         &       --         &        --        &        --       \\
\hline
&  spaxel (1,0) & spaxel (1,1) & spaxel (1,2) & spaxel (1,3) & spaxel (1,4)\\
\hline
$[\ion{O}{i}]\,63\,\mum$   &  $0.03\pm0.03$ & $0.99 \pm 0.09$ & $1.37 \pm 0.15$  & $0.60 \pm 0.11$  & $0.02 \pm 0.02$ \\
$[\ion{N}{ii}]\,122\,\mum$ &  $0.02\pm0.02$ & $0.15 \pm 0.08$ & $0.21 \pm 0.09$  & $0.14 \pm 0.07$  & $0.02 \pm 0.02$ \\
$[\ion{O}{i}]\,146\,\mum$  &  $0.01\pm0.01$ & $0.06 \pm 0.02$ & $0.08 \pm 0.02$  & $0.07 \pm 0.02$  &        --       \\
$[\ion{C}{ii}]\,158\,\mum$ &  $0.24\pm0.08$ & $0.27 \pm 0.07$ & $0.35 \pm 0.65$  & $0.29 \pm 0.06$  & $0.08 \pm 0.05$ \\
$[\ion{N}{ii}]\,205\,\mum$ &  $0.01\pm0.01$ &        --       & $0.01 \pm 0.01$  &  $0.01 \pm 0.01$ &        --       \\
\hline
&  spaxel (2,0) & spaxel (2,1) & spaxel (2,2) & spaxel (2,3) & spaxel (2,4)\\
\hline
$[\ion{O}{i}]\,63\,\mum$   &  $0.02\pm0.02$ & $1.22 \pm 0.21$ & $2.91 \pm 0.29$  & $2.44 \pm 0.30$  & $0.02 \pm 0.02$ \\
$[\ion{N}{ii}]\,122\,\mum$ &  $0.03\pm0.03$ & $0.15 \pm 0.05$ & $0.39 \pm 0.06$  & $0.27 \pm 0.07$  &        --       \\
$[\ion{O}{i}]\,146\,\mum$  &        --      & $0.39 \pm 0.07$ & $0.18 \pm 0.05$  & $0.18 \pm 0.05$  &        --       \\
$[\ion{C}{ii}]\,158\,\mum$ &  $0.21\pm0.09$ & $0.27 \pm 0.06$ & $0.45 \pm 0.05$  & $0.38 \pm 0.07$  & $0.10 \pm 0.05$ \\
$[\ion{N}{ii}]\,205\,\mum$ &        --      & $0.01 \pm 0.01$ & $0.02 \pm 0.01$  & $0.01 \pm 0.01$  &        --       \\
\hline
&  spaxel (3,0) & spaxel (3,1) & spaxel (3,2) & spaxel (3,3) & spaxel (3,4)\\
\hline
$[\ion{O}{i}]\,63\,\mum$   &  $0.02\pm0.02$ & $0.05 \pm 0.03$ & $1.48 \pm 0.36$  & $0.67 \pm 0.27$  & $0.01 \pm 0.01$ \\
$[\ion{N}{ii}]\,122\,\mum$ &  $0.01\pm0.01$ & $0.03 \pm 0.02$ & $0.24 \pm 0.06$  & $0.13 \pm 0.05$  &        --       \\
$[\ion{O}{i}]\,146\,\mum$  &         --     & $0.01 \pm 0.01$ & $0.13 \pm 0.07$  & $0.07 \pm 0.04$  &        --       \\
$[\ion{C}{ii}]\,158\,\mum$ &  $0.10\pm0.07$ & $0.16 \pm 0.06$ & $0.27 \pm 0.05$  & $0.21 \pm 0.06$  & $0.08 \pm 0.04$ \\
$[\ion{N}{ii}]\,205\,\mum$ &         --     &         --      & $0.01 \pm 0.01$  & $0.01 \pm 0.01$ &        --       \\
\hline
&  spaxel (4,0) & spaxel (4,1) & spaxel (4,2) & spaxel (4,3) & spaxel (4,4)\\
\hline
$[\ion{O}{i}]\,63\,\mum$   &  $0.01\pm0.01$ & $0.02 \pm 0.02$ & $0.02 \pm 0.02$  & $0.01 \pm 0.01$  & $0.02 \pm 0.02$ \\
$[\ion{N}{ii}]\,122\,\mum$ &  $0.02\pm0.02$ & $0.03 \pm 0.02$ & $0.05 \pm 0.03$  & $0.03 \pm 0.03$  &         --      \\
$[\ion{O}{i}]\,146\,\mum$  &          --    &         --      & $0.02 \pm 0.02$  &         --       & $0.01 \pm 0.01$ \\
$[\ion{C}{ii}]\,158\,\mum$ &  $0.08\pm0.05$ &         --      & $0.12 \pm 0.07$  & $0.13 \pm 0.07$  & $0.06 \pm 0.04$ \\
$[\ion{N}{ii}]\,205\,\mum$ &          --    &         --      &         --       &        --        & $0.01 \pm 0.01$ \\
\hline                                   
\end{tabular}
\end{table*}

\end{appendix}

\end{document}